\documentclass[final,12pt]{elsarticle}
\pdfoutput=1
\usepackage{silence} 
\usepackage{graphicx}
\usepackage[utf8]{inputenc} 
\usepackage{listings}
\usepackage{xargs}
\usepackage{xcolor} 
\usepackage{amsmath}
\usepackage{amsthm} 
\usepackage{algpseudocodex}
\usepackage{algorithmicx,algorithm, refcount}
\usepackage{comment}
\usepackage{caption}
\usepackage{subcaption}
\usepackage{mathdots}
\usepackage{yhmath}
\usepackage{cancel}
\usepackage{color}
\usepackage{siunitx}
\usepackage{array}
\usepackage{multirow}
\usepackage{amssymb}
\usepackage{gensymb}
\usepackage{tabularx}
\usepackage[big]{titlesec}
\usepackage{rotating}
\usepackage{xifthen}
\usepackage{mathtools}
\usepackage{enumitem}
\usepackage{stmaryrd} 
\usepackage{thmtools} 
\usepackage{varioref} 
\usepackage{hyperref} 
\usepackage[capitalise, noabbrev]{cleveref} 

\usepackage{tikz}
\usetikzlibrary{arrows.meta}
\usetikzlibrary{positioning}
\usetikzlibrary{calc}
\usetikzlibrary{fadings}
\usetikzlibrary{patterns}
\usetikzlibrary{shadows.blur}
\usetikzlibrary{backgrounds}
\usetikzlibrary{arrows}
\usetikzlibrary{shapes,shapes.geometric,shapes.misc}
\usetikzlibrary{decorations.pathmorphing}
\usetikzlibrary{decorations.pathreplacing}
\usetikzlibrary{decorations.markings}
\usetikzlibrary{math}

\tikzstyle{every picture}=[baseline=-0.25em]

\pgfkeys{/tikz/tikzit fill/.initial=0} \pgfkeys{/tikz/tikzit
draw/.initial=0} \pgfkeys{/tikz/tikzit shape/.initial=0}
\pgfkeys{/tikz/tikzit category/.initial=0}

\newcommand{\tikzfig}[1]{%
\IfFileExists{#1.tikz}
  {\input{#1.tikz}}
  {%
    \IfFileExists{./figures/#1.tikz}
      {\input{./figures/#1.tikz}}
      {\tikz[baseline=-0.5em]{\node[draw=red,font=\color{red},fill=red!10!white] {\textit{#1}};}}%
  }%
}

\pgfdeclarelayer{edgelayer} \pgfdeclarelayer{nodelayer}
\pgfsetlayers{background,edgelayer,nodelayer,main}
\tikzstyle{none}=[inner sep=0mm] \tikzstyle{every loop}=[]
\tikzstyle{mark coordinate}=[inner sep=0pt,outer sep=0pt,minimum
size=3pt,fill=black,circle]

\tikzstyle{nodo}=[fill=black, draw=black, shape=circle, inner
sep=1.5pt] \tikzstyle{class 1}=[fill=red, draw=red, shape=circle,
inner sep=1.5pt] \tikzstyle{class 2}=[fill=blue, draw=blue,
shape=circle, inner sep=1.5pt] \tikzstyle{class 3}=[fill=green,
draw=green, shape=circle, inner sep=1.5pt] \tikzstyle{class
4}=[fill=magenta, draw=magenta, shape=circle, inner sep=1.5pt]
\tikzstyle{class 5}=[fill={rgb,255: red,255; green,128; blue,0},
draw={rgb,255: red,255; green,128; blue,0}, shape=circle, inner
sep=1.5pt] \tikzstyle{class 6}=[fill=yellow, draw=yellow,
shape=circle, inner sep=1.5pt]

\tikzstyle{arista}=[-, fill=none, draw=black]

\newcommand{\declarefunction}[2][]{%
    \ifthenelse{\isempty{#1}}{%
        \expandafter\newcommand\csname #2\endcsname{\mathsf{#2}}%
    }{%
        \expandafter\newcommand\csname #1\endcsname{\mathsf{#2}}%
    }%
}
\newcommand{\declarevariable}[2][]{%
    \ifthenelse{\isempty{#1}}{%
        \expandafter\newcommand\csname #2\endcsname{\textup{\texttt{#2}}}%
    }{%
        \expandafter\newcommand\csname #1\endcsname{\textup{\texttt{#2}}}%
    }%
}

\declarefunction{thin}
\declarefunction{mimw}
\declarefunction{lmimw}
\declarefunction{pw}
\declarefunction{prev}
\declarefunction{child}
\declarefunction{gchild}
\declarefunction{type}
\declarefunction{thinList}
\declarefunction{critList}
\declarefunction{lastType}

\declarevariable{nil}
\declarevariable[childInfo]{child\_info}
\declarevariable[gchildInfo]{gchild\_info}

\DeclarePairedDelimiter\abs{\lvert}{\rvert}
\DeclarePairedDelimiter\dbrackets{\llbracket}{\rrbracket}
\DeclarePairedDelimiter\angles{\langle}{\rangle}
\DeclarePairedDelimiter\ceil{\lceil}{\rceil}

\renewcommand*{\emptyset}{\varnothing}
\newcommand*{\N}[3][]{N^{#1}_{#2}(#3)}
\newcommand*{\dangling}[3]{#1\langle #2,#3\rangle}
\newcommand*{\rooted}[2]{#1\dbrackets*{#2}}
\newcommand*{\bigO}{\textup{O}}
\newcommand*{\NN}{\mathbb{N}}
\newcommand*{\simplePath}{\mathcal{P}}
\newcommand*{\suffix}[2]{#1_{\geq #2}}

\newcommand*{\defref}[1]{\hyperref[#1]{definition of the \nameref*{#1}}}
\newcommand*{\subtreeWithoutCriticals}[2]{#1^{\setminus #2}}

\newtheorem{theorem}{Theorem}
\newtheorem{definition}[theorem]{Definition}

\newtheorem{lemma}[theorem]{Lemma}
\newtheorem{corollary}[theorem]{Corollary}

\newtheorem{observation}[theorem]{Observation}

\titlespacing*{\section}{0pt}{15pt}{15pt}
\titlespacing*{\subsection}{0pt}{10pt}{10pt}

\date{}
\begin{document}

\begin{frontmatter}
\title{On the thinness of trees}

\author[UBA,ICC]{Flavia Bonomo-Braberman}
\ead{fbonomo@dc.uba.ar}

\author[UBA]{Eric Brandwein}
\ead{ebrandwein@dc.uba.ar}

\author[UBA,ICC]{Carolina Luc\'{\i}a Gonzalez}
\ead{cgonzalez@dc.uba.ar}

\author[UBA]{Agust\'{\i}n Sansone}
\ead{asansone@dc.uba.ar}

\address[UBA]{Universidad de Buenos Aires. Facultad de Ciencias Exactas y Naturales. Departamento de Computaci\'on. Buenos Aires,
Argentina.}
\address[ICC]{CONICET - Universidad de Buenos Aires. Instituto de
Investigaci\'on en Ciencias de la Computaci\'on (ICC). Buenos
Aires, Argentina.}
\begin{abstract}

The study of structural graph width parameters like tree-width,
clique-width and rank-width has been ongoing during the last five
decades, and their algorithmic use has also been increasing
[Cygan~et~al., 2015]. New width parameters continue to be defined,
for example, MIM-width in 2012, twin-width in 2020, and
mixed-thinness, a generalization of thinness, in 2022.

The concept of \emph{thinness} of a graph was introduced in 2007
by Mannino, Oriolo, Ricci and Chandran, and it can be seen as a
generalization of interval graphs, which are exactly the graphs
with thinness equal to one. This concept is interesting because if
a representation of a graph as a $k$-thin graph is given for a
constant value $k$, then several known NP-complete problems can be
solved in polynomial time. Some examples are the maximum weighted
independent set problem, solved in the seminal paper by
Mannino~et~al., and the capacitated coloring with fixed number of
colors [Bonomo, Mattia and Oriolo, 2011].

In this work we present a constructive
$\bigO(n\log(n))$-time algorithm to compute the thinness for
any given $n$-vertex tree, along with a corresponding thin
representation. We use intermediate results of this construction
to improve known bounds of the thinness of some special families
of trees.

\nocite{cygan2015parameterized,M-O-R-C-thinness,B-M-O-thin-tcs,VatshelleThesis,BKTW20twinw,Mixed-thin-wg}

\end{abstract}
\end{frontmatter}

\section{Introduction}

A graph $G$ is \emph{$k$-thin} when it admits an order and a partition into $k$ classes on the vertices of $G$ satisfying that, for every ordered
triple $x < y < z$ such that $x$ and $y$ belong to the same class,
if $z$ is adjacent to $x$ then $z$ is also adjacent to $y$. In
that case, the order and partition are said to be
\textit{consistent}. The \emph{thinness} of a graph $G$ is the
minimum $k$ such that $G$ is $k$-thin.

The concept of thinness of a graph was introduced in 2007 by
Mannino, Oriolo, Ricci and Chandran~\cite{M-O-R-C-thinness}, and
it can be seen as a generalization of interval graphs, which are
exactly the graphs with thinness equal to one~\cite{Ola-interval,R-PR-interval}. This concept is
interesting because if a representation of a graph as a $k$-thin
graph is given for a constant value $k$, then several known
NP-complete problems can be solved in polynomial time. Some
examples are the maximum weighted independent set problem, solved
in the seminal paper by Mannino~et~al., and the capacitated
coloring with fixed number of colors~\cite{B-M-O-thin-tcs}. These
two examples can be generalized into a wide family of problems
that can be solved in XP parameterized by thinness, given a
consistent representation. This family includes weighted
variations of list matrix assignment problems with matrices of
bounded size, and the possibility of adding bounds on the weight
of the sets and their unions and
intersections~\cite{B-D-thinness}.

For a given order of the vertices of $G$, there exists an
algorithm to compute a consistent partition of the vertices of $G$
in the lowest amount of classes with time complexity
$\bigO(n^{3})$, where $n$ is the number of vertices of
$G$~\cite{B-D-thinness}, since the problem can be reduced in
linear time to the optimal coloring of an auxiliary
co-comparability graph of $n$ vertices, and the latter can be
solved in $\bigO(n^{3})$ time~\cite{Go-comp2}. On the other
hand, deciding the existence of an order on the vertices
consistent with a given partition is
NP-complete~\cite{B-D-thinness}. Very recently, by a reduction
from that problem, it was proved that deciding whether the
thinness of a graph is at most $k$, without any given order or
partition, is NP-complete~\cite{Shitov-thin}. In this work we
solve this problem in polynomial time for trees. This is the first
non-trivial efficient algorithm to compute the thinness (and
respective consistent order and partition of the vertices) within
a graph class.\footnote{A polynomial-time algorithm and forbidden
induced subgraphs characterization are known for thinness of
cographs~\cite{B-D-thinness}, but the algorithm and proofs follow
from the well known decomposition theorem of cographs without much
more complication.} Some efforts were made before to study the
thinness of trees in~\cite{Lucia-tesis}.

The design of this algorithm was heavily inspired by the proof and
the algorithm by H{\o}gemo, Telle, and V{\aa}gset for another
graph invariant, the \emph{linear maximum induced matching width}
(linear MIM-width)~\cite{H-A-R-lin-mim-trees}. The linear
MIM-width is a known lower bound for the
thinness~\cite{B-D-thinness}, and there are families with bounded
linear MIM-width and unbounded
thinness~\cite{B-B-M-P-convex-wads}. However, we prove here that,
for trees, the two parameters behave alike and the difference
between thinness and linear MIM-width is at most~1.

An extended abstract of this paper was presented at ISCO~2022 and
appears in~\cite{Agus-Eric-ISCO22}.

\section{Definitions and preliminary results}

\subparagraph*{Lists.}
We will use the operator $\oplus$ to add an element at the beginning of a list. For example, $1 \oplus (2,3) = (1,2,3)$. 

Let $L$ be a list, and $i \in \{1,\dots,\abs{L}\}$. We will use $L_i$ to refer to the element at position $i$ of $L$. For example, if $L = (1,2,3)$, then $L_2 = 2$.

Let $j \in \NN$. Let $L = (l_1,\dots,l_{\abs{L}})$. The \emph{suffix of $L$ starting at position $j$} is the list $(l_j,\dots,l_{\abs{L}})$, or the empty list if $j > \abs{L}$. We denote it by $\suffix{L}{j}$.
For example, if $L = (1,2,3)$, then $\suffix{L}{2} = (2,3)$.

Throughout the paper we use the operator $\in$ for membership in lists as well as sets. In particular, $x \in L$ means that $x$ is equal to an element of $L$. 

A common data structure in the algorithms we present is a \emph{linked list}, which allows adding and removing elements in constant time, and concatenating two lists also in constant time. A description of this data structure can be found in~\cite{art-of-computer-programming}.

\subparagraph*{Graphs.}
All graphs in this work are finite, undirected and have no loops
or multiple edges. 
Let $G$ be a graph, we denote by $V(G)$ its vertex set and by
$E(G)$ its edge set. We denote by $N(v)$ and $N[v]$, respectively,
the neighborhood and closed neighborhood of a vertex $v \in V(G)$.
Let $X \subseteq V(G)$. We denote by $N(X)$ the set of vertices
not in $X$ having at least one neighbor in $X$, and by $N[X]$ the
closed neighborhood $N(X) \cup X$.

We denote by $G[X]$ the subgraph of $G$ induced by $X$, and by $G \setminus W$ the graph $G[V(G) \setminus W]$.
We use $G \setminus (u,v)$ to denote the graph with vertices
$V(G)$ and edges $E(G) \setminus \{(u,v)\}$. If $H$ is also a graph, we use $G \cup H$ to denote the graph $(V(G) \cup V(H), E(G) \cup E(H))$, and $G \cap H$ to denote the graph $(V(G) \cap V(H), E(G) \cap E(H))$.

\subparagraph*{Trees.}
A \emph{tree} is a connected graph with no cycles. A \emph{leaf}
of a tree $T$ is a vertex with degree one in $T$. The
\emph{diameter} of a tree is the maximum number of edges in a
simple path joining two vertices. If $u$ and $v$ are two vertices in a tree $T$, then we denote by $\simplePath_T(u,v)$ the simple path that connects $u$ and $v$ in $T$, or simply $\simplePath(u,v)$ if $T$ is clear from the context.

A \emph{rooted tree on vertex $r$} is a tree in which vertex $r$
is labeled as the root, and we will usually denote it by $T_r$.
The \emph{ancestors} of a vertex $v$ in a rooted tree $T_r$ are
all vertices in the simple path between $v$ and $r$ which
are not $v$. Note that $r$ has no ancestors in $T_r$. The
\emph{descendants} of a vertex $v$ in $T_r$ are all vertices for
which $v$ is an ancestor in $T_r$. The \emph{children} of a vertex
$v$ are those neighbors of $v$ which are also descendants, and we denote them $\child(v)$.
Conversely, the \emph{parent} of a vertex $v$ different from $r$ is the only
neighbor of $v$ which is also an ancestor of $v$. In a
rooted tree $T_r$, the vertex $r$ has no parent. The
\emph{grandchildren} of a vertex $v$ are the children's children, denoted $\gchild(v)$,
and the \emph{grandparent} is the parent's parent, if any exists. A \emph{subtree} of a tree $T$ is an induced subgraph of $T$. A \emph{supertree} of a tree $T$ is a tree $T'$ such that $T$ is a subtree of $T'$. A \emph{strict subtree} of a tree $T$ is a subtree of $T$ which is different from $T$.

The \emph{height} of a rooted tree is the maximum number of edges
in a simple path from the root to a leaf. A rooted tree of height
$h$ has $h+1$ \emph{levels} of vertices, where the root is the only
vertex at level 1, and a vertex has level $t+1$ if and
only if its parent has level $t$.

An $m$-ary tree is a rooted tree in which all vertices but the leaves
have $m$ children. A \textit{complete} $m$-ary tree is an $m$-ary
tree in which all leaves are at the same distance from the root.

Let $T$ be a tree containing the adjacent vertices $v$ and $u$.
The \emph{dangling tree} from $v$ in $u$, denoted $\dangling{T}{v}{u}$, is
the component of $T \setminus (v, u)$ containing $u$.

\subparagraph*{Other graph classes and graph parameters.}
A graph $G$ is \emph{$k$-thin} if there exists a strict total order $\prec$ on $V(G)$ and a partition $S$ of $V(G)$ into
$k$ classes such that, for each $u, v, w \in V(G)$ with $u \prec v \prec
w$, if $u$ and $v$ belong to the same class and $(u,w) \in
E(G)$, then $(v,w) \in E(G)$. An order and a partition
satisfying those properties are said to be \textit{consistent} with $G$. We
call the tuple $(\prec, S)$ a \emph{consistent solution} for $G$. The minimum $k$ such that $G$ is
$k$-thin is called the \textit{thinness} of $G$, and denoted by
$\thin(G)$. A consistent solution for $G$ that uses $\thin(G)$ classes is said to be \emph{optimal}.

A class of graphs $\mathcal{C}$ is \emph{hereditary} if every induced subgraph of a graph $G \in \mathcal{C}$ is also in $\mathcal{C}$. The following result is used throughout the literature with no proof, and we include it here for completeness.
\begin{lemma}[Heredity Lemma]
    \label{lemma:hereditary}
    Let $G$ be a graph. For every induced subgraph $H$ of $G$, $\thin(H) \leq \thin(G)$. In other words, the class of $k$-thin graphs is hereditary for every $k$.
\end{lemma}
\begin{proof}
    Let $(S, \prec)$ be an optimal consistent solution for $G$. Let $S_H$ be the partition of $V(H)$ induced by $S$, and $\prec_H$ the order on $V(H)$ induced by $\prec$. The pair $(S_H, \prec_H)$ is a consistent solution for $H$ using at most $\thin(G)$ classes, and then $\thin(H) \leq \thin(G)$.
\end{proof}

A \textit{complete graph} is a graph where all vertices are
pairwise adjacent. We denote by $K_n$ the complete graph of $n$
vertices. A \emph{clique} of a graph $G$ is a complete induced
subgraph of $G$, and the \emph{clique number} of $G$ is the
maximum size (number of vertices) of a clique of $G$. An
\emph{independent set} of a graph is a set of pairwise
non-adjacent vertices. A \textit{coloring} of a graph is a
partition of its vertex set into independent sets called
\emph{color classes}. The minimum number of classes in such a
partition is called the \emph{chromatic number} of the graph. A
graph is \emph{perfect} whenever for every induced subgraph of it,
the chromatic number equals the clique number.

An \emph{interval representation} of a graph $G$ is a family of
closed intervals $\{I_v\}_{v \in V(G)}$ on the real line such that
two distinct vertices $u,v \in V(G)$ are adjacent if and only
if $I_u \cap I_v \neq \emptyset$. A graph $G$ is an \emph{interval
graph} if it admits an interval representation. 
Graphs of thinness one are exactly the
interval graphs~\cite{Ola-interval}.

A graph $G$ is a \emph{comparability graph} if there exists a
partial order in $V(G)$ such that two vertices of $G$ are adjacent
if and only if they are comparable by that order. A graph $G$ is a
\emph{co-comparability graph} if its complement $\overline{G}$ is
a comparability graph.

Let $G$ be a graph and $\prec$ a strict total order on $V(G)$. The graph
$G_\prec$ has $V(G)$ as vertex set, and $E(G_\prec)$ is such that for every two vertices
$v,w \in V(G)$ such that $v \prec w$, edge $(v,w)$ is in $E(G_\prec)$ if and only if there is a
vertex $z$ in $G$ such that:
\begin{itemize}
    \item $w \prec z$;
    \item $(z, v) \in E(G)$;
    \item $(z,w) \notin E(G)$.
\end{itemize}
An edge of $G_\prec$ represents that its endpoints
cannot belong to the same class in a vertex partition that is
consistent with $\prec$.

\begin{theorem}[\cite{B-D-thinness}] Given a graph $G$ and an order $\prec$ on its vertices, the following statements are true:
    \begin{itemize}
        \item $G_\prec$ is a co-comparability graph;
        \item the chromatic number of $G_\prec$ is equal to the minimum integer $k$ such that there is a partition of $V(G)$ into $k$ sets that is consistent with the order $\prec$;
        \item the color classes of a valid coloring of $G_\prec$ form a partition consistent with $\prec$.
    \end{itemize}
\end{theorem}

Since co-comparability graphs are perfect~\cite{Meyn-co-comp}, in
$G_\prec$ the chromatic number equals the clique number. We thus have
the following.
\begin{corollary}
    \label{coro:G<-clique}
    Let $G$ be a graph, and $k$ a positive integer. Then $\thin(G) \geq k$
if and only if, for every order $\prec$ on $V(G)$, the graph $G_\prec$
has a clique of size $k$.
\end{corollary}

A graph $G'$ is a \emph{supergraph} of a graph $G$ if $G$ is a subgraph of $G'$.
An \emph{interval supergraph} of $G$ is a supergraph of $G$ that is an interval graph.
The \textit{pathwidth} of a graph $G$, denoted by $\pw(G)$, is the
minimum clique number of all interval supergraphs of $G$, minus
one~\cite{K-S-bw}. 

Let $G$ be a graph, and let $\sigma$ be a strict total order on $V(G)$ such that $v_1 < \dots < v_n$ according to $\sigma$. For every $i \in \{1,\dots,n\}$ we denote by $G_i$ the graph induced by the edges in $G$ with one endpoint in $\{v_1,\dots,v_i\}$ and the other one in $\{v_{i+1},\dots,v_n\}$.
The \emph{maximum induced matching width of $G$ under $\sigma$}, denoted $\mimw(\sigma, G)$, is the value of the maximum induced matching of all $G_i$. The \textit{linear MIM-width} of a graph $G$,
denoted by $\lmimw(G)$, is the minimum value of $\mimw(\sigma, G)$ of all possible orders $\sigma$ of $V(G)$~\cite{VatshelleThesis}.

The following relations are known.

\begin{theorem}\label{lmimw-pw-bounds}\cite{B-D-thinness,M-O-R-C-thinness} For every graph $G$, $\lmimw(G) \leq \thin(G) \leq \pw(G)+1$. \end{theorem}

\section[Characterization and Algorithm]{Structural characterization of the thinness of trees}
A considerable part of the ideas related to the construction of
the algorithm we present in this section to compute the thinness (and a consistent order
and partition of the vertices) was inspired by an algorithm to compute the linear MIM-width of a
tree and an optimal layout~\cite{H-A-R-lin-mim-trees}, which was
at the same time inspired by the framework behind the pathwidth
algorithm presented in~\cite{E-S-T-pw-trees}.

In this section, we characterize the thinness of trees in terms of vertices we call \emph{$k$-saturated vertices}. We show that a tree has thinness at least $k+1$ if and only if it has a $k$-saturated vertex. First, we present a lemma that allows us to construct a consistent solution starting from a particular path of a tree, and then we use this lemma to prove the characterization.
Finally, we present a polynomial time algorithm to find these $k$-saturated vertices and compute the thinness of a tree with a corresponding consistent solution.

\subsection{Path Layout Lemma}

\begin{lemma}[Path Layout Lemma]\label{lem:path-layout}
Let $T$ be a tree. If there exists a path $P$ in
$T$ such that every connected component of $T \setminus N[P]$ has
thinness lower than or equal to $k$, then $\thin(T) \leq k + 1$.
Moreover, if $q$ is the number of connected components in $T \setminus N[P]$,
then, given the consistent orders and partitions for the
components of $T \setminus N[P]$ in at most $k$ classes, we can in
$\bigO(\abs{V(N[P])} + k \cdot q)$ compute a consistent order and partition for $T$ in
at most $k+1$ classes.
\end{lemma}

\begin{proof}
We give \vref{consistent-layout}, which constructs a solution $(\sigma_T, S_T)$ with $k+1$ classes for $T$ using the consistent solutions with $k$ classes for the connected components of $T \setminus N[P]$, and show that the returned solution is consistent, implying that \(\thin(T) \leq k + 1 \).

The algorithm returns an order $\sigma_T$ of the vertices of $T$, and a partition $S_T$ of $V(T)$ into $k+1$ classes. Adding vertex $w_1$ before vertex $w_2$ to $\sigma_T$ is interpreted as $w_1 < w_2$ in the returned order, and adding vertex $w$ to a class is interpreted as $w$ belonging to that class in the returned partition.

In the algorithm, $p$ is the length of the path $P$, and $(x_1,\dots,x_p)$ correspond to the vertices in $P$. For $i \in \{1,\dots,p\}$ and $j \in \{1,\dots,|N(x_i) \setminus P|\}$,
vertex $v_{i,j}$ corresponds to a neighbor of $x_i$ which is not in $P$. For $m \in \{1,\dots,|N(v_{i,j}) \setminus \{ x_i \}|\}$, vertex $u_{i,j,m}$ corresponds to a neighbor of $v_{i, j}$ 
different from $x_i$. The connected components of $T \setminus N[P]$ are exactly the dangling trees 
$\dangling{T}{v_{i,j}}{u_{i,j,m}}$, which we denote $T_{i,j,m}$, and so the parameters of this algorithm 
are the consistent orders and partitions in $k$ classes for these trees. For every $c \in \{1,\dots,k + 1\}$,
we use \(C^T_c\) $\in \NN$ to denote
class number \(c\) of \(S_T\), and for every $q \in \{1,\dots,k\}$, we use
\(C^{T_{i,j,m}}_q\) to denote class number \(q\)
of \(S_{T_{i,j,m}}\).

Each order $\sigma_{T_{i,j,m}}$ is represented by a linked list of vertices. Each partition $S_{T_{i,j,m}}$ is represented by a list of linked lists. Each linked list corresponds to a class in the partition, and each entry in each linked list is a vertex belonging to that class. The returned order $\sigma_T$ is also represented by a linked list of vertices, and the returned partition $S_T$ is represented by a list where each class $C_c^T$ is a linked list of vertices.

\begin{algorithm}
\caption{Compute a consistent solution for $T$ given the solutions for the dangling
trees of a path in $T$.} \label{consistent-layout}
    \begin{algorithmic}[1]
        \Function{ConsistentSolutionGivenPath}{%
            \\\phantom{\textbf{function}}\(T\): tree,
            \\\phantom{\textbf{function}}\( P = (x_1,\dots,x_p) \) : path,
            \\\phantom{\textbf{function}}\( \{\sigma_{T_{i,j,m}}\}_{\forall T_{i,j,m}} \): orders,
            \\\phantom{\textbf{function}}\( \{ S_{T_{i,j,m}} = \{C^{T_{i,j,m}}_1,\dots,C^{T_{i,j,m}}_k\}\}_{\forall T_{i,j,m}} \): partitions%
        }
            \State Assign the empty linked list to $\sigma_T$
            \For{\( c \in \{1,\dots,k+1\} \)}
                \State Assign the empty linked list to class $C^T_c$ of $S_T$
            \EndFor
            \For{\( x_i \in P \)}
                \For{\( v_{i,j} \in N(x_i) \setminus P\)}
                    \State Append $v_{i,j}$ to $\sigma_T$ and add it to class $C^T_{k+1}$ \label{alg:consistent:append-vij}
                    \For{\( u_{i,j,m} \in N(v_{i,j}) \setminus \{x_i\}\)}
                        \State Concatenate $\sigma_{T_{i,j,m}}$ to the end of $\sigma_T$ \label{alg:consistent:append-uijm}
                        \For{\( c \in \{1,\dots,k\} \)}
                            \State Merge $C^{T_{i,j,m}}_c$ into $C^T_c$ \label{alg:consistent:add-uijm}
                        \EndFor
                    \EndFor
                \EndFor
                \State Append $x_i$ to $\sigma_T$ and add it to class $C^T_{k+1}$ \label{alg:consistent:append-xi}
            \EndFor
        \EndFunction
    \end{algorithmic}
\end{algorithm}

For each $i \in \{1,\dots,p\}$ and $j \in \{1,\dots,|N(x_i) \setminus P|\}$, \Cref{consistent-layout} adds vertex $x_i \in P$ to $\sigma_T$ after every neighbor $v_{i,j}$. For each $m \in \{1,\dots,|N(v_{i,j}) \setminus \{ x_i \}|\}$, every vertex in $T_{i,j,m}$ is added right before its respective neighbor $v_{i,j}$ in $N[P]$. The returned solution maintains the order and partition of the solutions for the dangling trees, thus maintaining consistency. Also, it assigns class $C^T_{k+1}$ to all vertices in $N[P]$, and uses classes $C^T_1,\dots,C^T_k$ for the vertices in the dangling trees.

First, we show that every vertex in $N[P]$ is appended exactly once to $\sigma_T$ and added to a single class of $S_T$ in \cref{consistent-layout}. Indeed, every vertex $x_i \in P$ is processed exactly once (\cref{alg:consistent:append-xi}), and every neighbor $v_{i,j}$ of $x_i$ not in $P$ is processed exactly once (\cref{alg:consistent:append-vij}). As appending to a linked list is a constant time operation, this operations cost in total $\bigO(\abs{V(N[P])})$.

Second, we note that every vertex in every dangling tree appears exactly once in $\sigma_T$ (\cref{alg:consistent:append-uijm}) and belongs to a single class of $S_T$ (\cref{alg:consistent:add-uijm}) in \cref{consistent-layout}. Thus, $\sigma_T$ is a valid order on $V(T)$, and $S_T$ is a valid partition of $V(T)$. Merging two linked lists takes constant time, so both operations in \cref{alg:consistent:append-uijm} and \cref{alg:consistent:add-uijm} have constant time complexity. \cref{alg:consistent:append-uijm} is executed a number of times proportional to the number $q$ of connected components of $T \setminus N[P]$. Meanwhile, \cref{alg:consistent:add-uijm} is executed a number of times proportional to $k \cdot q$. Thus, these two lines incur a runtime complexity of $\bigO(k \cdot q)$ in total. 

These are the only operations performed by the algorithm apart from the initialization of the solution, which has time complexity $\bigO(k)$. Thus, in total the algorithm has time complexity $\bigO(\abs{V(N[P])} + k \cdot q)$.

Now we show that the
order and partition are consistent, meaning, there are no three
vertices $w_1, w_2, w_3$ in \( \sigma_T\) such that \( w_1 < w_2 < w_3 \), vertices \( w_1 \) and \(
w_2 \) are in the same class of the partition, and \( (w_1,w_3) \in E(T)
\) and \( (w_2,w_3) \not\in E(T) \). To do this, we analyze each possible triple of vertices $\{w_1, w_2, w_3\} \subseteq V(T)$ such that \( w_1 < w_2 < w_3 \) and see that it does not violate consistency.
\begin{itemize}
    \item \emph{%
        $w_1 \in N[P]$ and $w_2 \not\in N[P]$, or $w_1 \not\in N[P]$ and $w_2 \in N[P]$%
    }:
        All vertices in $N[P]$ are added to class $C^T_{k+1}$, and all vertices not in $N[P]$ are added to classes $C^T_1,\dots,C^T_k$. Thus, $w_1$ and $w_2$ cannot be in the same class in $S_T$, and so this triple is consistent.
        
    \item \emph{\( w_1 \not\in N[P] \text{ and } w_3 \in N[P]\)}:
        Vertex $w_3$ is not adjacent to $w_1$, as the only vertices
        of \( N[P] \) adjacent to vertices not in \( N[P] \) are the \( v_{i,j}\)
        for some $i \in \{1,\dots,p\}$ and $j \in \{1,\dots,|N(x_i) \setminus P|\}$, and they are added to the order before all their neighbors.
        Thus, this triple is consistent.

    \item \emph{\( w_1 \in N[P] \text{ and } w_3 \not\in N[P]\)}:
        As before, vertex \( w_1 \) must be equal to \( v_{i,j} \) for some $i \in \{1,\dots,p\}$ and $j \in \{1,\dots,|N(x_i) \setminus P|\}$
        to be adjacent to a vertex \( w_3 \not\in N[P] \). Also, the only vertices not
        in \( N[P] \) adjacent to \( v_{i,j} \) are the \( u_{i,j,m} \) for some $m \in \{1,\dots,|N(v_{i,j}) \setminus \{ x_i \}|\}$, so
        $w_3$ either is not adjacent to $w_1$, and thus this triple is consistent, or $w_3 = u_{i,j,m}$.
        The only possible vertices between \( v_{i,j} \) and \( u_{i,j,m} \) in $\sigma_T$ are
        the vertices of \( T_{i,j,m} \), which means that
        \( w_2 \in T_{i,j,m} \). But then \( w_1 \in N[P] \)
        and \( w_2 \not\in N[P] \), which means they are in different classes of the partition,
        and so this triple is consistent.
\end{itemize}
Combining the last three cases we see that, to violate consistency, either the three vertices
must belong to \( N[P] \), or none of the three can.
\begin{itemize}
    \item \emph{\( \{w_1,w_2,w_3\} \subseteq N[P]\)}:
        We begin by noting that, for $i \in \{1,\dots,p\}$ and $j \in \{1,\dots,|N(x_i) \setminus P|\}$, no vertex \( v_{i,j} \) is adjacent to another vertex \( y \)
        in \( N[P] \) such that \( y < v_{i,j} \), because \( v_{i,j} \) is always
        added to the order before \( x_i \), which is the only vertex in
        \( N[P] \) adjacent to \( v_{i,j} \). This means that for $w_3$ to be adjacent to $w_1$, it must be equal to some $x_i$. The only vertices adjacent to \( x_i \) that are preceding in $\sigma_T$
        are all \( v_{i,j} \), and \( x_{i-1}\) if \( i > 1\).

        If \( w_1 = v_{i,j} \) for some $i \in \{1,\dots,p\}$ and $j \in \{1,\dots,|N(x_i) \setminus P|\}$, then all vertices \( w_2 \in N[P] \) such that
        \( w_1 < w_2 < x_i \) are \( v_{i,k} \) for some \( k > j\).
        As \( (x_i, v_{i,k}) \in T \), this triple is consistent.
        If, on the other hand, \( w_1 = x_{i-1} \), then again, \( w_2 = v_{i,k} \)
        for some $k \in \{1,\dots,|N(x_i) \setminus P|\}$, which means it is also adjacent to \( x_i \), and so this triple is
        also consistent.

    \item \emph{%
        \( \{w_1,w_2,w_3\} \subseteq V(T_{i,j,m})\) for some $i \in \{1,\dots,p\}$, $j \in \{1,\dots,|N(x_i) \setminus P|\}$, and $m \in \{1,\dots,|N(v_{i,j}) \setminus \{ x_i \}|\}$%
    }:
        Because \( \sigma_{T_{i,j,m}}\) is a subsequence of \(\sigma_T\),
        and all classes of \( S_{T_{i,j,m}}\) are subsets of the
        corresponding classes of \( S_T \), the consistency is preserved between three vertices of
        the same dangling tree.

    \item \emph{%
        \( w_1 \in V(T_{i,j,m})\) and
        \(w_3 \in V(T_{i',j',m'})\) for some $i, i' \in \{1,\dots,p\}$, $j \in \{1,\dots,|N(x_i) \setminus P|\}$, $j' \in \{1,\dots,|N(x_{i'}) \setminus P|\}$, $m \in \{1,\dots,|N(v_{i,j}) \setminus \{ x_i \}|\}$, and $m' \in \{1,\dots,|N(v_{i',j'}) \setminus \{ x_{i'} \}|\}$ such that \( i \neq i',j \neq j', \text{ or } m \neq m' \)%
    }:
        Vertices in different dangling trees are not adjacent, so this triple is consistent.

    \item \emph{%
        \( \{w_1,w_3\} \subseteq V(T_{i,j,m})\) and 
        \(w_2 \in V(T_{i',j',m'})\) for some $i, i' \in \{1,\dots,p\}$, $j \in \{1,\dots,|N(x_i) \setminus P|\}$, $j' \in \{1,\dots,|N(x_{i'}) \setminus P|\}$, $m \in \{1,\dots,|N(v_{i,j}) \setminus \{ x_i \}|\}$, and $m' \in \{1,\dots,|N(v_{i',j'}) \setminus \{ x_{i'} \}|\}$ such that \( i \neq i',j \neq j', \text{ or } m \neq m' \)%
    }:
        Either all the vertices of \( T_{i,j,m} \) are added to $\sigma_T$ before all
        the vertices of \(T_{i',j',m'}\), or vice versa. This means that
        either \( w_1 < w_2 \) and \( w_3 < w_2 \), or \( w_2 < w_1 \) and \( w_2 < w_3 \), so this case cannot happen.
\end{itemize}

We have proved that any possible triple in the order and partition
generated by the algorithm is consistent, and then the order and
partition given by the algorithm are consistent.

\end{proof}

\subsection{Structural characterization}

We will utilize the \nameref{lem:path-layout} to prove the characterization of the thinness of trees in terms of $k$-saturated vertices. We first define the notion of $k$-saturated vertices.

\begin{definition}[$k$-neighbor]
    \label{def:k-neighbor}
    Let $X$ be a subset of vertices of a tree $T$, and $v$ a vertex in $N(X)$.
    If there exists a neighbor $u$ of $v$ not in $X$ such that $\thin(\dangling{T}{v}{u}) \geq k$, then $v$ is a \emph{$k$-neighbor of $X$}. If $X$ contains only one vertex $w$, we also say that $v$ is a $k$-neighbor of $w$.
\end{definition}

\begin{definition}[$k$-neighborhood]
    \label{def:k-neighborhood}
    Let $T$ be a tree, and let $k \in \NN$. For each subset of vertices $X$ of $V(T)$ we use $\N[T]{k}{X}$ to denote the \emph{$k$-neighborhood of $X$}, meaning, the set of $k$-neighbors of $X$ in $T$. If $X$ contains only one vertex $w$, then $\N[T]{k}{w}$ denotes the set of $k$-neighbors of $w$ in $T$. If $T$ is clear from the context, we simply write $\N{k}{X}$ or $\N{k}{w}$.
\end{definition}

\begin{definition}[$k$-saturation]
    \label{def:k-saturation}
    Let $X$ be a subset of vertices of a tree $T$, and let $k \in \NN$.
    We say that $X$ is \emph{$k$-saturated in $T$} if $\abs*{\N[T]{k}{X}} \geq 3$. If $T$ is clear from the context, we simply say that $X$ is $k$-saturated. If $X$ contains only one vertex $w$, we also say that $w$ is $k$-saturated. 
\end{definition}

Now, we proceed with the characterization. We will show that a tree has thinness at least $k+1$ if and only if it has a $k$-saturated vertex. The proof has two parts: in \cref{lem1-mainth} we prove the backward implication, and in \cref{lem2-mainth} the forward implication.

\newcommand{\first}{x}
\newcommand{\last}{z}
\begin{lemma}\label{lem1-mainth}
    Let $k \in \NN$. If \( \abs*{\N{k}{w}} \geq 3 \) for some vertex \( w \) in \( T \), then \( \thin(T) \geq k + 1 \).
\end{lemma}

\begin{proof}
    Let $w$ be a vertex in $T$ such that \( \abs*{\N{k}{w}} \geq 3 \).
    Let \( v_1, v_2\), and \( v_3 \) be three neighbors of \( w \) that have neighbors
    \( u_1, u_2\), and \(u_3 \) respectively which satisfy that \( \thin(\dangling{T}{v_i}{u_i}) \geq k\) for each $i \in \{1,2,3\}$. Denote \( T_i = \dangling{T}{v_i}{u_i} \).
    Let $ \prec $ be an order on the vertices of $T$ which is part of an optimal consistent solution for $T$.
    Let $\first$ and $\last$ be, respectively, the lowest and greatest vertex according to $\prec$ in $\bigcup_{i\in\{1,2,3\}}V(T_i)$.

    By the pigeonhole principle, there must be at least one subtree \( T_j \) for some \(j\in \{1,2,3\} \) such that \( \first \not\in T_j \) and
    \( \last \not\in T_j \). As \( \thin(T_j) \geq k \), by \cref{coro:G<-clique} we know that \( T_{j_\prec}\) has a clique
    \( C \) of size at least \( k \). Let \( y \) be the greatest vertex according to \( \prec \) that belongs
    to \( C \).

    We know that \( \first \prec y \prec \last \), because \( \first \) is lower in the order than all the vertices in
    \( T_j \), and \( \last \) is greater. Also, there is a simple path $P$ between \( \first \) and \( \last \)
    that does not include any vertex of \( T_j \), nor any vertex adjacent to any other vertex in $T_j$. In other words, $P$ and $N[V(T_j)]$ are disjoint sets. Indeed,
    if \( \first \) and \( \last \) belong to the same subtree \( T_i \) for some $ i\in \{1, 2, 3\} $ different from $j$, then there is a simple
    path \( P \) between the two that includes only vertices of \( T_i \). The only neighbor of any vertex in $T_i$ that is not in $T_i$ is $v_i$, which is not adjacent to any vertex in $T_j$. 
    On the other hand, if \( \first \) and \( \last \) belong to different subtrees \( T_h \) and \( T_i \) for some $h, i \in \{1,2,3\}$ different from $j$, then
    the simple path is
    \( \first \rightarrow \dots \rightarrow u_h \rightarrow v_h \rightarrow w \rightarrow v_i
    \rightarrow u_i \rightarrow \dots \rightarrow \last\),
    which does not include any vertex of \( N[V(T_j)] \).

    Since \( P \) begins with a vertex which is lower than \( y \) in $\prec$, and ends with
    a vertex greater than \( y \), there exist two adjacent vertices \( x', z' \in P \) such that
    \( x' \prec y \prec z' \). We will see that
    \( x' \) is adjacent in \( T_\prec \) to all the vertices in \( C \), which means that there is a
    clique of size \( k + 1 \) in \( T_\prec \).

    We know that
    \( x' \prec y \prec z'\), and \( (x', z') \in E(T) \), but, as every vertex in $P$ is not adjacent to any vertex in $T_j$, \( (y, z') \not\in E(T) \). This, by definition of $T_\prec$, means that \( x' \) is adjacent to \( y \) in \( T_\prec \).

    Now, given $y'$, a vertex of \( C \) such that $y' \prec y$, let us see that \( (x', y') \in E(T_\prec) \).
    As \( y' \) is adjacent to \( y \) in \( T_{j_\prec} \), there is a vertex $y^*$ in \( T_j \) that forms an inconsistent triple with $y'$ and $y$, meaning, it holds that $y' \prec y \prec y^*$, $y'$ is adjacent to $y^*$, and $y$ is not adjacent to $y^*$. As \( x' \prec y \prec y^* \), it follows that \( x' \prec y^* \). Also, from \( y' \prec y \prec z' \) follows that \( y' \prec z' \). This leaves us with two possibilities:
    \begin{itemize}
        \item \( x' \prec y' \): vertex \( z' \) is adjacent in \( T \) to \( x' \) but not to
            \( y' \), so \( x' \) and \( y' \) are adjacent in \( T_\prec \).
        \item \( y' \prec x' \): vertex \( y^* \) is adjacent in \( T \) to
        \( y' \) but not to \( x' \), so \( x' \) and \( y' \) are adjacent in \( T_\prec \).
    \end{itemize}

    This shows that every vertex of \( C \) is adjacent to a vertex
    \( x' \) in \( T_\prec \), and so \( T_\prec \) has a clique of size at least \( k + 1 \),
    which, by \cref{coro:G<-clique}, implies that \( \thin(T) \geq k+1 \).
\end{proof}

\begin{lemma}\label{lem2-mainth}
    Let $k \in \NN$. If $\thin(T) \geq k + 1$, then there exists a vertex $x$ in $T$ such that $\abs*{\N{k}{x}} \geq 3$.
\end{lemma}

\begin{proof}
We prove the contrapositive statement, so let us assume that
every vertex $x$ in $T$ is such that $\abs*{\N{k}{x}} < 3$ and show that then $\thin(T
) \leq k$. We show that there is always a path $P$ in $T$
such that all the connected components in $T \setminus N[P]$ have
thinness lower than or equal to $k-1$. This way, using the \nameref{lem:path-layout}, we
show that $\thin(T) \leq k$.

We begin by defining the following two sets of vertices:
\[X = \{x\ |\ x \in V(T)\ \textrm{and}\ \abs*{\N{k}{x}} = 2\}\]
\[Y = \{y\ |\ y \in V(T)\ \textrm{and}\ \abs*{\N{k}{y}} = 1\}\]

\emph{\underline{Case 1: $X \neq \emptyset$}}
\medskip

\newcommand{\Q}{Q}
First, we claim that every pair of vertices in $X$ are connected by a path in $X$. If $X$ has only one vertex, or $X$ has exacly two vertices which are neighbors, this is trivial. Suppose then that $X$ has at least two distinct vertices that are not neighbors, $x_i$ and $x_j$. Take the simple path $\Q=(x_i , \dots , x_j)$ connecting $x_i$ and $x_j$. 

It suffices to show that each element $x_l$ of $\Q$ different from $x_i$ and $x_j$ has two $k$-neighbors, and thus $x_l \in X$. To see this, first note that $x_i$ and
$x_j$ each have at least one $k$-neighbor not belonging to $\Q$, as they each
have at most one neighbor in $\Q$, and $\abs*{\N{k}{x_i}} = \abs*{\N{k}{x_j}} = 2$.
Let $x_i'$ and $x_j'$ be $k$-neighbors of $x_i$ and $x_j$ that do not belong to $\Q$, respectively. We extend the path $\Q$ with these two $k$-neighbors to obtain $\Q'$, defined as $(x_i', x_i, \dots, x_j, x_j')$.

Let $x_{l-1}$ and $x_{l-2}$ be the two vertices that come right before $x_l$ in $\Q'$.  Symmetrically, let $x_{l+1}$ and $x_{l+2}$ be the two vertices that come right after $x_l$ in $\Q'$. These exist because $x_l \not\in \{x_i', x_i, x_j, x_j'\}$. 

Let $x_i''$ be a neighbor of $x_i'$ different from $x_i$ such that $\thin(\dangling{T}{x_i'}{x_i''}) \geq k$.
The dangling tree $\dangling{T}{x_{l-1}}{x_{l-2}}$ has $\dangling{T}{x_i'}{x_i''}$ as an induced subgraph, and so by the \nameref{lemma:hereditary}, $\thin(\dangling{T}{x_{l-1}}{x_{l-2}}) \geq k$. This implies that $x_{l-1}$ is a $k$-neighbor of $x_l$. An analogous argument can be used to prove that $x_{l+1}$ is a $k$-neighbor of $x_l$, and thus $x_l$ has two $k$-neighbors. This proves that $x_l \in X$, and thus our claim is true.

Notice that in this proof, vertex $x_{l-1}$ is just a neighbor of $x_l$ in a path to $x_l$ from an arbitrary vertex $x_i$ in $X$. In particular, every neighbor of $x_l$ that belongs to $X$ also belongs to a path between $x_l$ and a vertex in $X$. This means that every neighbor of $x_l$ that belongs to $X$ is a $k$-neighbor of $x_l$.

The fact that every pair of vertices in $X$ are connected by a
path in $X$ means that $X$ must be a connected subtree of $T$. Furthermore, this subtree must be a path. Otherwise, there would be a vertex $w \in X$ of degree at least 3 in $T[X]$, which would mean that $w$ has at least 3 neighbors in $X$, which must be $k$-neighbors of $w$. This cannot happen, as $\abs*{\N{k}{w}} = 2$.

We therefore conclude that all vertices in $X$ must lie on some
path $\Q=(x_1 , \dots , x_p)$ of size $p \in \NN$. Let us see now that we can apply the
\nameref{lem:path-layout}.

Let $x_0$ be the $k$-neighbor of $x_1$ which is not in $X$, and
$x_{p+1}$ be the $k$-neighbor of $x_p$ which is not in $X$. If
there is only one vertex $x$ in $X$, then let $x_0$ and $x_{p+1}$
be the two $k$-neighbors of $x$. Vertices $x_0$ and $x_{p+1}$ only
have one \(k\)-neighbor ($x_1$ and $x_p$ respectively) or else
they would be in $X$. Every other $k$-neighbor of a vertex in $X$
is itself in $X$. 

Let $P = (x_0, \dots , x_{p+1})$. Every connected component in $T \setminus N[P]$ must
have thinness not greater than $k - 1$, as no vertex $v \in N(P) \setminus P$ is a
$k$-neighbor of a vertex in $P$. We can then apply the \nameref{lem:path-layout} with $P$ to $T$ to obtain that $\thin(T) \leq k$.

\medskip
\emph{\underline{Case 2: $X = \emptyset, Y \neq \emptyset$}}
\medskip

We will construct the path $P$ which will be used to apply the
\nameref{lem:path-layout}. We start with $P=(y_1 , y_2)$, where $y_1$ is
some arbitrary vertex in $Y$, and $y_2$ its only \(k\)-neighbor.
Then, if the last vertex in $P$ has a \(k\)-neighbor $y' \notin
P$, we append $y'$ to $P$, and repeat this process exhaustively.
Since we are only considering finite graphs, we will eventually reach
some vertex $y_p$ for some $p \in \NN$ such that either $y_p \notin Y$ or the
\(k\)-neighbor of $y_p$ is in $P$. We are then done and have
$P=(y_1, \dots , y_p)$, which is a path in $T$ by construction.

One property of $P$ is that no vertex $y_i \in P$ for some $i \in \{1,\dots,p\}$ can have a
$k$-neighbor not belonging to $P$. In the case of $i = p$, this is by
construction. In the case of $i \neq p$, if $y_i$ had a
$k$-neighbor outside $P$, it would have at least two $k$-neighbors
(the other one being $y_{i+1}$) which cannot happen because $X$ is
empty. This means that $T \setminus N[P]$ has no subtrees with
thinness greater than $k-1$. We can then apply the \nameref{lem:path-layout} with $P$ to $T$ to obtain that $\thin(T) \leq k$.

\medskip
\emph{\underline{Case 3: $X = \emptyset, Y = \emptyset$}}
\medskip

Let $v$ be an arbitrary vertex in $T$, and $P$ be the path with only $v$ as an element.
As both $X$ and $Y$ are empty, $v$ has no $k$-neighbors. Thus, no
subtree of $T \setminus N[P]$ has thinness greater than $k-1$. We can then apply the \nameref{lem:path-layout} with $P$ to $T$ to obtain that $\thin(T) \leq k$.

\end{proof}

These two lemmas can be synthesized in the following theorem.

\begin{theorem}[Characterization Theorem]
\label{thm:characterization}
    Let $k \in \NN$. The thinness of a tree $T$ is greater than $k$ if and only if $\abs*{\N{k}{x}} \geq 3$ for some vertex $x \in V(T)$.
\end{theorem}

\begin{proof} 
    It follows directly from \cref{lem1-mainth,lem2-mainth}.
\end{proof}

The \nameref{thm:characterization} can be extended to paths in the following way.

\begin{corollary}
\label{coro:characterization-by-paths}
    Let $k \in \NN$. The thinness of a tree $T$ is greater than $k$ if and only if there is a $k$-saturated path in $T$.
\end{corollary}
\begin{proof}
    $\Rightarrow$) If $\thin(T) \geq k$, then by the \nameref{thm:characterization} it contains a $k$-saturated vertex $x$. If we take $P$ to be the path in $T$ containing only $x$, we see that $P$ is a $k$-saturated path in $T$.

    $\Leftarrow$) Let $P$ be a path in $T$ such that $\abs*{\N{k}{P}} \geq 3$. Let $v_1$, $v_2$ and $v_3$ be three distinct $k$-neighbors of $P$. For every $i \in \{1,2,3\}$, let $x_i$ be the neighbor of $v_i$ in $P$, and let $u_i$ be a neighbor of $v_i$ different from $x_i$ such that $\dangling{T}{v_i}{u_i}$ has thinness at least $k$. Without loss of generality, assume that $x_2$ does not appear before $x_1$ in $P$, and $x_3$ does not appear before $x_2$. \Cref{fig:characterization-by-paths} shows a diagram of this situation.

    \begin{figure}[htbp]
        \centering
        \begin{tikzpicture}
            \node (dots1) {\dots};    
            \foreach \x in {1,...,3} {
                \pgfmathtruncatemacro{\next}{\x + 1}
                \node[right=of dots\x] (x\x) {$x_\x$};
                \node[below=1cm of x\x] (v\x) {$v_\x$};
                \node[below=1cm of v\x] (u\x) {$u_\x$};
                
                \node[below=2cm of u\x] (base) {};
                \node[below=1cm of u\x] (tree) {\footnotesize{$\dangling{T}{v_\x}{u_\x}$}};
                \node[right=of base] (right-angle) {};
                \node[left=of base] (left-angle) {};
                
                \node[right=of x\x] (dots\next) {\dots};
                
                \draw (dots\x) -- (x\x) -- (dots\next);
                \draw (x\x) -- (v\x) -- (u\x);
                \draw (u\x) -- (left-angle.center) -- (right-angle.center) -- (u\x);
            };

            \node[above left=0.15cm of dots1] (P-left) {};
            \node[below right=0.15cm of dots4] (P-right) {};
            \draw[rounded corners=4mm, draw=blue, fill=blue, fill opacity=0.1] (P-left) rectangle (P-right);
            \node[above=0.5cm of dots1] {\Large\color{blue} $P$};

            \coordinate[below left=0.4cm of u1] (Pp1);
            \coordinate[below right=0.4cm of u1] (Pp2);
            \coordinate[below right=0.4cm of x1] (Pp3);
            \coordinate[below left=0.4cm of x3] (Pp4);
            \coordinate[below left=0.4cm of u3] (Pp5);
            \coordinate[below right=0.4cm of u3] (Pp6);
            \coordinate[above right=0.4cm of x3] (Pp7);
            \coordinate[above left=0.4cm of x1] (Pp8);

            \draw[rounded corners=6mm, draw=red, fill=red, fill opacity=0.1]
                (Pp1)
                \foreach \i in {2,...,8} { -- (Pp\i) }
                -- cycle;
            
            \node[above right=0.6cm of u3] {\Large\color{red}$P'$};
        \end{tikzpicture}
        \caption{Paths $P$ and $P'$ in tree $T$.}
        \label{fig:characterization-by-paths}
    \end{figure}
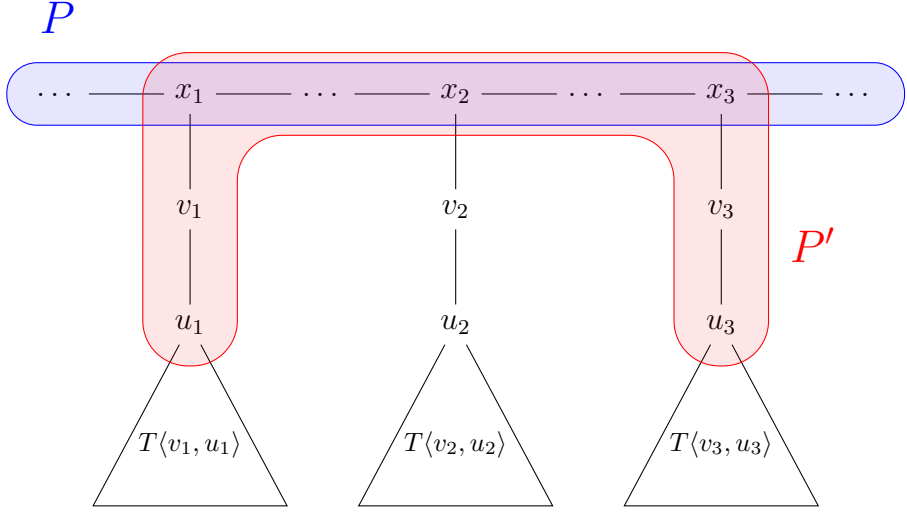

    We will show that $x_2$ is $k$-saturated, and hence $T$ has thinness greater than $k$ by the \nameref{thm:characterization}.

    Denote by $P'$ the path $\simplePath(u_1, u_3)$ as seen in \cref{fig:characterization-by-paths}. This path contains all the vertices in $\{v_1, x_1, x_2, x_3, v_3\}$, because $T$ is a tree. Let $w_1$ and $w_2$ be the two vertices that appear right before $x_2$ in $P'$, and $y_1$ and $y_2$ be the two vertices that appear right after $x_2$. The dangling tree $\dangling{T}{v_1}{u_1}$ is a subtree of $\dangling{T}{w_2}{w_1}$, so by the \nameref{lemma:hereditary} $\dangling{T}{w_2}{w_1}$ has thinness at least $k$. This means that $w_2$ is a $k$-neighbor of $x_2$. An analogous argument can be used to prove that $y_1$ is a $k$-neighbor of $x_2$. As $v_2$ is also a $k$-neighbor of $x_2$, vertex $x_2$ is $k$-saturated, and so $T$ has thinness greater than $k$ by the \nameref{thm:characterization}.
\end{proof}

H{\o}gemo et al.~\cite{H-A-R-lin-mim-trees} proved a similar result to the \nameref{thm:characterization}   for linear MIM-width. They define the $k$-component index for linear MIM-width as follows.

\begin{definition}[Linear MIM-width $k$-neighbor and $k$-component index {\cite[Definition~2]{H-A-R-lin-mim-trees}}]
    Let $x$ be a vertex in the tree $T$ and $v$ a neighbor of $x$. If $v$ has a neighbor $u \neq x$ such that $\lmimw(\dangling{T}{v}{u}) \geq k$ for some $k \in \NN$, then we call $v$ a \emph{linear MIM-width $k$-neighbor of $x$}. The \emph{linear MIM-width $k$-component index of $x$} is equal to the number of linear MIM-width $k$-neighbors of $x$ and is denoted $D_T(x, k)$, or shortened to $D(x,k)$.
\end{definition}

\begin{theorem}[Classification of the Linear MIM-width of Trees {\cite[Theorem~1]{H-A-R-lin-mim-trees}}]
\label{k-neighbors-bound-lmimw}
    Let $T$ be a tree and $k\geq 1$, then $\lmimw(T) \geq k+1$ if and only if $D(x, k) \geq 3$ for some vertex $x$ in $T$.
\end{theorem}

We can then prove the following.

\begin{corollary}\label{cor:lmimw-vs-thin}
    For any given tree $T$, $\thin(T)-\lmimw(T) \leq 1$.
\end{corollary}

\begin{proof}
    We will prove by induction on $k$ that for every tree $T$ with $\thin(T) \geq k$, $\lmimw(T) \geq k - 1$.

    \begin{itemize}
        \item $k = 1$: The minimum linear MIM-width of a graph is 0, so $\lmimw(T) \geq k - 1$.
        \item $k = 2$: Trees with thinness greater or equal to 2 have at least two connected vertices, and so they contain $K_2$ as an induced subgraph, which has linear MIM-width~1. The linear MIM-width of $T$ is then at least 1, as the linear MIM-width is a hereditary property.
        \item $k > 2$: We take as inductive hypothesis that every tree with thinness greater or equal to $k-1$ has linear MIM-width greater or equal to $k-2$. In this case, by the \nameref{thm:characterization}, $T$ contains a vertex $x$ such that $\abs*{\N{k-1}{x}} \geq 3$. This means there are at least three
        subtrees $T_1$, $T_2$, and $T_3$ dangling from neighbors of $x$ that each have thinness greater or equal to $k-1$. By the inductive hypothesis, these subtrees have linear MIM-width greater or equal to $k-2$. This means that $D(x, k-2) \geq 3$. As $k-2$ is greater than 0, we can use \cref{k-neighbors-bound-lmimw} to prove that $\lmimw(T) \geq k-1$.
    \end{itemize}
\end{proof}

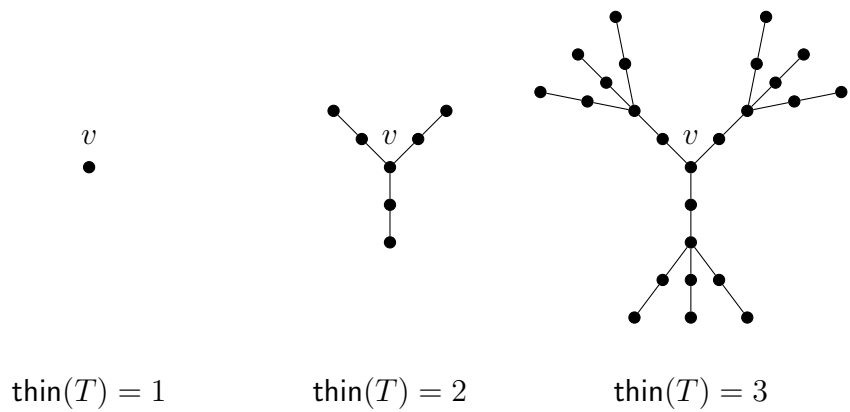
\begin{figure}[htbp]
\begin{center}
\begin{tikzpicture}[scale=0.5]
    \begin{pgfonlayer}{nodelayer}
        \node [style=nodo] (0) at (0, 0) {};
        \node [style=nodo] (1) at (-0.75, 0.75) {};
        \node [style=nodo] (2) at (0, -1) {};
        \node [style=nodo] (3) at (0.75, 0.75) {};
        \node [style=nodo] (4) at (-1.5, 1.5) {};
        \node [style=nodo] (5) at (0, -2) {};
        \node [style=nodo] (6) at (1.5, 1.5) {};
        \node [style=nodo] (7) at (-8, 0) {};
        \node [style=nodo] (8) at (8, 0) {};
        \node [style=nodo] (9) at (7.25, 0.75) {};
        \node [style=nodo] (10) at (8, -1) {};
        \node [style=nodo] (11) at (8.75, 0.75) {};
        \node [style=nodo] (12) at (6.5, 1.5) {};
        \node [style=nodo] (13) at (8, -2) {};
        \node [style=nodo] (15) at (6.25, 2.75) {};
        \node [style=nodo] (16) at (6, 4) {};
        \node [style=nodo] (17) at (5.75, 2.25) {};
        \node [style=nodo] (18) at (5, 3) {};
        \node [style=nodo] (19) at (5.25, 1.75) {};
        \node [style=nodo] (20) at (4, 2) {};
        \node [style=nodo] (22) at (9.5, 1.5) {};
        \node [style=nodo] (23) at (10.75, 1.75) {};
        \node [style=nodo] (24) at (10.25, 2.25) {};
        \node [style=nodo] (25) at (11, 3) {};
        \node [style=nodo] (26) at (9.75, 2.75) {};
        \node [style=nodo] (27) at (10, 4) {};
        \node [style=nodo] (28) at (12, 2) {};
        \node [style=nodo] (29) at (7.25, -3) {};
        \node [style=nodo] (30) at (8.75, -3) {};
        \node [style=nodo] (31) at (6.5, -4) {};
        \node [style=nodo] (32) at (8, -3) {};
        \node [style=nodo] (33) at (9.5, -4) {};
        \node [style=nodo] (34) at (8, -4) {};
        \node [style=none] (35) at (-8, -6) {};
        \node [style=none] (36) at (-8, -6) {};
        \node [style=none] (37) at (-8, -6) {};
        \node [style=none] (38) at (-8, -6) {$\thin(T) = 1$};
        \node [style=none] (39) at (0, -6) {$\thin(T) = 2$};
        \node [style=none] (40) at (8, -6) {$\thin(T) = 3$};
        \node [style=none] (41) at (-8, 0.8) {$v$};
        \node [style=none] (42) at (0, 0.8) {$v$};
        \node [style=none] (43) at (8, 0.8) {$v$};
    \end{pgfonlayer}
    \begin{pgfonlayer}{edgelayer}
        \draw [style=arista] (0) to (3);
        \draw [style=arista] (3) to (6);
        \draw [style=arista] (0) to (2);
        \draw [style=arista] (2) to (5);
        \draw [style=arista] (0) to (1);
        \draw [style=arista] (1) to (4);
        \draw [style=arista] (8) to (11);
        \draw [style=arista] (8) to (10);
        \draw [style=arista] (10) to (13);
        \draw [style=arista] (8) to (9);
        \draw [style=arista] (9) to (12);
        \draw [style=arista] (12) to (15);
        \draw [style=arista] (15) to (16);
        \draw [style=arista] (12) to (19);
        \draw [style=arista] (12) to (17);
        \draw [style=arista] (17) to (18);
        \draw [style=arista] (19) to (20);
        \draw [style=arista] (22) to (23);
        \draw [style=arista] (22) to (26);
        \draw [style=arista] (22) to (24);
        \draw [style=arista] (24) to (25);
        \draw [style=arista] (26) to (27);
        \draw (23) to (28);
        \draw (11) to (22);
        \draw (13) to (29);
        \draw (29) to (31);
        \draw (13) to (32);
        \draw (13) to (30);
        \draw (30) to (33);
        \draw (32) to (34);
    \end{pgfonlayer}
\end{tikzpicture}
\end{center}
\caption{Smallest trees for each thinness
value.}\label{fig:thinness}
\end{figure}

The difference arises from the fact that every graph has thinness
at least one, while edgeless graphs have linear MIM-width zero.
Indeed, we can use the \nameref{thm:characterization} to find
the smallest trees for each thinness value. For each thinness value
$k$, the vertex $v$ in \vref{fig:thinness} is such that
$\abs*{\N{k-1}{v}} = 3$. The smallest tree with thinness $k$ can be
constructed by replacing each leaf in the smallest tree with
thinness~$2$ with one of the smallest trees with thinness $k-1$, thus
achieving $\abs*{\N{k}{v}} = 3$ with the minimum amount of vertices. 

Note that to construct a smallest tree of thinness $k > 1$, the neighbors of $v$ can be adjacent to any of the vertices of a smallest tree with thinness $k-1$. Thus, the trees of thinness 1 and 2 pictured in \cref{fig:thinness} are unique, while the trees of thinness 3 and above are not.  

\begin{figure}[htbp]
\begin{center}
\begin{tikzpicture}[scale=0.5]
    \begin{pgfonlayer}{nodelayer}
        \node [style=nodo] (0) at (0, 0) {};
        \node [style=nodo] (1) at (-0.75, 0.75) {};
        \node [style=nodo] (2) at (0, -1) {};
        \node [style=nodo] (3) at (0.75, 0.75) {};
        \node [style=nodo] (4) at (-1.5, 1.5) {};
        \node [style=nodo] (5) at (0, -2) {};
        \node [style=nodo] (6) at (1.5, 1.5) {};
        \node [style=nodo] (7) at (-8, 0) {};
        \node [style=nodo] (8) at (8, 0) {};
        \node [style=nodo] (9) at (7.25, 0.75) {};
        \node [style=nodo] (10) at (8, -1) {};
        \node [style=nodo] (11) at (8.75, 0.75) {};
        \node [style=nodo] (12) at (6.5, 1.5) {};
        \node [style=nodo] (13) at (8, -2) {};
        \node [style=nodo] (15) at (6.5, 2.5) {};
        \node [style=nodo] (16) at (6, 3) {};
        \node [style=nodo] (17) at (6, 2) {};
        \node [style=nodo] (18) at (5.5, 2.5) {};
        \node [style=nodo] (19) at (5.5, 1.5) {};
        \node [style=nodo] (20) at (5, 2) {};
        \node [style=nodo] (29) at (7.25, -2.75) {};
        \node [style=nodo] (30) at (8.75, -2.75) {};
        \node [style=nodo] (31) at (7.25, -3.5) {};
        \node [style=nodo] (32) at (8, -2.75) {};
        \node [style=nodo] (33) at (8.75, -3.5) {};
        \node [style=nodo] (34) at (8, -3.5) {};
        \node [style=none] (35) at (-8, -6) {};
        \node [style=none] (36) at (-8, -6) {};
        \node [style=none] (37) at (-8, -6) {};
        \node [style=none] (38) at (-8, -6) {$\lmimw(T) = 1$};
        \node [style=none] (39) at (0, -6) {$\lmimw(T) = 2$};
        \node [style=none] (40) at (8, -6) {$\lmimw(T) = 3$};
        \node [style=none] (41) at (-8, 0.8) {$v$};
        \node [style=none] (42) at (0, 0.8) {$v$};
        \node [style=none] (43) at (8, 0.8) {$v$};
        \node [style=nodo] (44) at (-8, -1) {};
        \node [style=nodo] (45) at (0, -3) {};
        \node [style=nodo] (46) at (-2.25, 2.25) {};
        \node [style=nodo] (47) at (2.25, 2.25) {};
        \node [style=nodo] (48) at (4.5, 2.5) {};
        \node [style=nodo] (49) at (5, 3) {};
        \node [style=nodo] (50) at (5.5, 3.5) {};
        \node [style=nodo] (51) at (9.5, 1.5) {};
        \node [style=nodo] (52) at (10.5, 1.5) {};
        \node [style=nodo] (53) at (11, 2) {};
        \node [style=nodo] (54) at (10, 2) {};
        \node [style=nodo] (55) at (10.5, 2.5) {};
        \node [style=nodo] (56) at (9.5, 2.5) {};
        \node [style=nodo] (57) at (10, 3) {};
        \node [style=nodo] (58) at (10.5, 3.5) {};
        \node [style=nodo] (59) at (11, 3) {};
        \node [style=nodo] (60) at (11.5, 2.5) {};
        \node [style=nodo] (61) at (7.25, -4.25) {};
        \node [style=nodo] (62) at (8, -4.25) {};
        \node [style=nodo] (63) at (8.75, -4.25) {};
    \end{pgfonlayer}
    \begin{pgfonlayer}{edgelayer}
        \draw [style=arista] (0) to (3);
        \draw [style=arista] (3) to (6);
        \draw [style=arista] (0) to (2);
        \draw [style=arista] (2) to (5);
        \draw [style=arista] (0) to (1);
        \draw [style=arista] (1) to (4);
        \draw [style=arista] (8) to (11);
        \draw [style=arista] (8) to (10);
        \draw [style=arista] (10) to (13);
        \draw [style=arista] (8) to (9);
        \draw [style=arista] (9) to (12);
        \draw [style=arista] (12) to (15);
        \draw [style=arista] (15) to (16);
        \draw [style=arista] (12) to (19);
        \draw [style=arista] (12) to (17);
        \draw [style=arista] (17) to (18);
        \draw [style=arista] (19) to (20);
        \draw (13) to (29);
        \draw (29) to (31);
        \draw (13) to (32);
        \draw (13) to (30);
        \draw (30) to (33);
        \draw (32) to (34);
        \draw (7) to (44);
        \draw (47) to (6);
        \draw (46) to (4);
        \draw (5) to (45);
        \draw (48) to (20);
        \draw (49) to (18);
        \draw (50) to (16);
        \draw [style=arista] (51) to (52);
        \draw [style=arista] (52) to (53);
        \draw [style=arista] (51) to (56);
        \draw [style=arista] (51) to (54);
        \draw [style=arista] (54) to (55);
        \draw [style=arista] (56) to (57);
        \draw (58) to (57);
        \draw (59) to (55);
        \draw (60) to (53);
        \draw (11) to (51);
        \draw (31) to (61);
        \draw (34) to (62);
        \draw (33) to (63);
    \end{pgfonlayer}
\end{tikzpicture}
\end{center}
\caption{Smallest trees for each linear MIM-width
value.}\label{fig:lmim-width}
\end{figure}
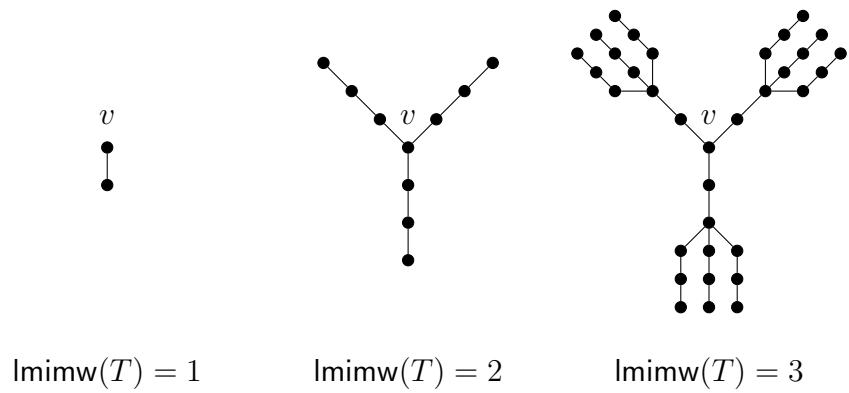

Compare this with the smallest trees with linear MIM-width 1, 2 and
3~\cite{Hogemo-thesis}, depicted in \cref{fig:lmim-width}.
These are pretty similar, except that the leaves in the trees with
thinness $k$ are replaced by two vertices. This is because the
\cref{k-neighbors-bound-lmimw} for linear MIM-width is very similar to the \nameref{thm:characterization}, and the
smallest tree with linear MIM-width 1 is the path of two vertices,
while for thinness 1 it is a single vertex. This produces slightly
bigger trees than for the thinness, which corresponds to the
fact that the linear MIM-width is a lower bound for the thinness.

\begin{figure}[htbp]
\begin{center}
\begin{tikzpicture}[scale=0.5]
    \begin{pgfonlayer}{nodelayer}
        \node [style=nodo] (0) at (0, 0) {};
        \node [style=nodo] (1) at (-0.75, 0.75) {};
        \node [style=nodo] (2) at (0, -1) {};
        \node [style=nodo] (3) at (0.75, 0.75) {};
        \node [style=nodo] (4) at (-1.5, 1.5) {};
        \node [style=nodo] (5) at (0, -2) {};
        \node [style=nodo] (6) at (1.5, 1.5) {};
        \node [style=nodo] (7) at (-8, 0) {};
        \node [style=nodo] (77) at (-8, -1) {};
        \node [style=nodo] (8) at (8, 0) {};
        \node [style=nodo] (9) at (7.25, 0.75) {};
        \node [style=nodo] (22) at (8.75, 0.75) {};
        \node [style=nodo] (12) at (6.5, 1.5) {};
        \node [style=nodo] (13) at (8, -1) {};
        \node [style=nodo] (15) at (5.75, 2.25) {};
        \node [style=nodo] (17) at (7, 2) {};
        \node [style=nodo] (18) at (6.75, 3.25) {};
        \node [style=nodo] (19) at (6, 1) {};
        \node [style=nodo] (20) at (4.75, 1.25) {};
        \node [style=nodo] (23) at (10, 1) {};
        \node [style=nodo] (24) at (9.5, 1.5) {};
        \node [style=nodo] (25) at (10.25, 2.25) {};
        \node [style=nodo] (26) at (9, 2) {};
        \node [style=nodo] (27) at (9.25, 3.25) {};
        \node [style=nodo] (28) at (11.25, 1.25) {};
        \node [style=nodo] (29) at (7.25, -2) {};
        \node [style=nodo] (30) at (8.75, -2) {};
        \node [style=nodo] (31) at (6.5, -3) {};
        \node [style=nodo] (32) at (8, -2) {};
        \node [style=nodo] (33) at (9.5, -3) {};
        \node [style=nodo] (34) at (8, -3) {};
        \node [style=none] (35) at (-8, -5) {};
        \node [style=none] (36) at (-8, -5) {};
        \node [style=none] (37) at (-8, -5) {};
        \node [style=none] (38) at (-8, -5) {$\pw(T) = 1$};
        \node [style=none] (39) at (0, -5) {$\pw(T) = 2$};
        \node [style=none] (40) at (8, -5) {$\pw(T) = 3$};
        \node [style=none] (41) at (-8, 0.8) {$v$};
        \node [style=none] (42) at (0, 0.8) {$v$};
        \node [style=none] (43) at (8, 0.8) {$v$};
    \end{pgfonlayer}
    \begin{pgfonlayer}{edgelayer}
        \draw [style=arista] (7) to (77);
        \draw [style=arista] (0) to (3);
        \draw [style=arista] (3) to (6);
        \draw [style=arista] (0) to (2);
        \draw [style=arista] (2) to (5);
        \draw [style=arista] (0) to (1);
        \draw [style=arista] (1) to (4);
        \draw [style=arista] (8) to (22);
        \draw [style=arista] (8) to (13);
        \draw [style=arista] (8) to (9);
        \draw [style=arista] (9) to (12);
        \draw [style=arista] (12) to (15);
        \draw [style=arista] (9) to (19);
        \draw [style=arista] (9) to (17);
        \draw [style=arista] (17) to (18);
        \draw [style=arista] (19) to (20);
        \draw [style=arista] (22) to (23);
        \draw [style=arista] (22) to (26);
        \draw [style=arista] (22) to (24);
        \draw [style=arista] (24) to (25);
        \draw [style=arista] (26) to (27);
        \draw [style=arista] (23) to (28);
        \draw [style=arista] (13) to (29);
        \draw [style=arista] (29) to (31);
        \draw [style=arista] (13) to (32);
        \draw [style=arista] (13) to (30);
        \draw [style=arista] (30) to (33);
        \draw [style=arista] (32) to (34);
    \end{pgfonlayer}
\end{tikzpicture}
\end{center}
\caption{Smallest trees for each pathwidth
value.}\label{fig:pathwidth}
\end{figure}
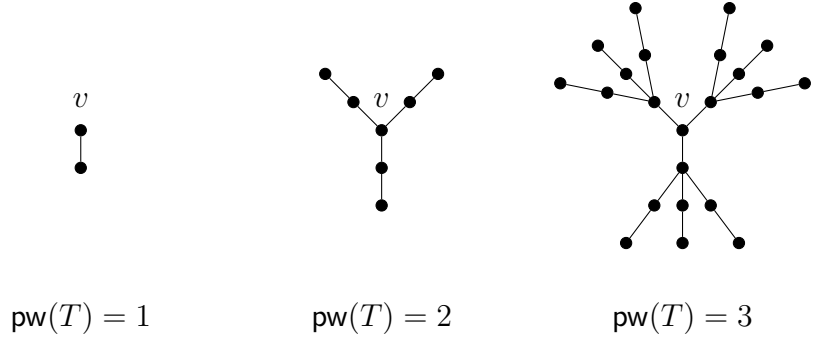

Regarding the pathwidth, instead, the smallest trees are smaller
(see \vref{fig:pathwidth}), which also corresponds to the
fact that the pathwidth plus one is an upper bound for the
thinness~\cite{M-O-R-C-thinness}. Again, a theorem similar to
the \nameref{thm:characterization} holds for
pathwidth~\cite{E-S-T-pw-trees}, but with subtrees instead of
dangling trees.

\subsection{Consequences of the characterization theorem}

\begin{corollary}[Bound on the number of vertices]
    \label{log-bound}
    The thinness of an $n$-vertex tree $T$ is $\bigO(\log(n))$.
    In fact $\thin(T) \leq \log_3(n+2)$.
\end{corollary}

\begin{proof}
    We will prove the statement by induction on the number of vertices $n$ of $T$.
    \begin{itemize}
        \item \emph{Base case} ($n = 1$): The tree with only one vertex has thinness 1, which satisfies $\thin(T) \leq \log_3(1+2) = 1$.
        \item \emph{Inductive step} ($n > 1$): Suppose that the property holds for all trees with strictly less than $n$ vertices.
        
        Let $k = \thin(T)$. If $k = 1$, then $\thin(T) \leq \log_3(n+2)$. Otherwise, by the \nameref{thm:characterization} there is a vertex $x \in V(T)$ such that $\abs*{\N{k-1}{x}} \geq 3$.

        Let $v_1, v_2$ and $v_3$ be three $k$-neighbors of $x$ such that $\dangling{T}{v_1}{u_1}$, $\dangling{T}{v_2}{u_2}$, and $\dangling{T}{v_3}{u_3}$ are pairwise disjoint dangling subtrees of $T$ with thinness at least $k-1$. Let $i \in \{1,2,3\}$, let $T_i = \dangling{T}{v_i}{u_i}$, and let $n_i = \abs{V(T_i)}$. By the inductive hypothesis, $k-1 \leq \log_3(n_i+2)$. This means that $T_i$ has at least $3^{k-1}-2$ vertices. This in turn implies that $T$ has at least $3(3^{k-1}-2)+4$ vertices when we account for $x$, $v_j$, and $T_j$, for all $j \in \{1,2,3\}$. As $3(3^{k-1}-2)+4 = 3^k-2$, the statement holds for $T$. 
    \end{itemize}
\end{proof}

\begin{corollary}[Bound on the number of leaves]
    \label{log-leaves-bound} A nontrivial tree of thinness at least $k$ has at least
    $\frac{3^{k-1}+3}{2}$ leaves. In particular, the thinness of a tree with $\ell$ leaves is at most $\log_3(6\ell-9)$.
\end{corollary}

\begin{proof}
    We will prove the statement by induction on the number of vertices $n$ of $T$.

    \begin{itemize}
        \item \emph{Base case} ($n = 2$): The tree with only two vertices has thinness 1 and two leaves, which makes $T$ satisfy the statement.
        \item \emph{Inductive hypothesis} ($n > 2$): Suppose that all trees with strictly less than $n$ vertices satisfy the statement.
        
        Let $k = \thin(T)$. If $k = 1$, then $T$ has at least $\frac{3^{k-1}+3}{2}$ leaves, as all nontrivial trees have at least two leaves. Otherwise, by the \nameref{thm:characterization} there is a vertex $x \in V(T)$ such that $\abs*{\N{k-1}{x}} \geq 3$.

        Let $v_1, v_2$ and $v_3$ be three $k$-neighbors of $x$ such that $\dangling{T}{v_1}{u_1}$, $\dangling{T}{v_2}{u_2}$, and $\dangling{T}{v_3}{u_3}$ are pairwise disjoint dangling subtrees of $T$ with thinness at least $k-1$. Let $i \in \{1,2,3\}$, let $T_i = \dangling{T}{v_i}{u_i}$, and let $n_i = \abs{V(T_i)}$. By the inductive hypothesis, $T_i$ has at least $\frac{3^{k-2}+3}{2}$ leaves. Since $u_i$ is the only vertex in $\dangling{T}{v_i}{u_i}$ that could be a leaf in $\dangling{T}{v_i}{u_i}$ but not in $T$, we can ensure
        that $T$ has at least $3\frac{3^{k-2}+3}{2}-3 = \frac{3^{k-1}+3}{2}$ leaves, and so $T$ satisfies the statement.
    \end{itemize}
\end{proof}

\cref{log-leaves-bound} establishes an upper bound on
the thinness of a graph in terms of the number of leaves. Let us
call an \emph{almost-leaf} a vertex which is not a leaf, that has
at most one neighbor that is not a leaf. Using some ideas
from~\cite{B-B-M-P-convex-jcss}, we can prove the following bound,
useful for trees with a big number of leaves but few
internal vertices of degree greater than two after trimming the
leaves.

\begin{theorem}
    Let $T$ be a tree with $t$ almost-leaves, where $t \geq 2$. Then $\thin(T) \leq t-1$.
\end{theorem}

\begin{proof}
We start by removing all the leaves of $T$, obtaining a tree $T'$.
The leaves of $T'$ are the almost-leaves of $T$. We will construct a partition of the vertices of $T'$, and then extend it by adding the leaves of $T$. 

We first root $T'$ at a leaf $r$ to obtain $T'_r$. Let $v_1,\dots, v_{t-1}$ be the leaves in the lowest level of $T'_r$. We create a new class $C_{v_i}$ in the partition for each $v_i$ with $i \in \{1, \dots, t-1\}$, and assign $v_i$ to $C_{v_i}$. Then, if a vertex $v$ has only one child $u$, then $v$ is assigned to the same
class as $u$. If $v$ has more than one child, then $v$ is assigned to any one of the classes of its children. Clearly, this partition has $t-1$ classes.

Now, we define an order $\prec$ on $V(T'_r)$ in which every vertex is smaller than its parent. Vertices that share a parent are ordered arbitrarily. We add each leaf $x \in T$ to $\prec$ right before its only adjacent vertex $v \in T'_r$, and assign it to the same class as $v$. Vertex $v$ exists, since $T'$ is not empty.

Note that the induced subgraph $T[C_{v_i}]$ is a subtree of $T$ for every $v_i \in \{v_1,\dots,v_{t-1}\}$, as every vertex in $C_{v_i}$ is connected to $v_i$ through a path in $C_{v_i}$.

We will now show that the order and the partition are consistent.
Suppose that $x \prec y \prec z$, with $x$ and $y$ in the same class of the partition
and $(x, z) \in E(T)$. We will show that $(y, z) \in E(T)$, proving that every triple of vertices is consistent.

First, notice that $z$ cannot be a leaf in $T$,
since if a vertex is a leaf, then its
only neighbor is greater than it. So $z$ is a vertex of $T'$. 

If $x$ is a leaf in $T$, then its only neighbor is $z$. All the leaves of $T$ adjacent to $z$ were added right before $z$ in the order. Thus $y$ must also be a leaf of $T$ adjacent to $z$ to satisfy $x \prec y \prec z$, and therefore $(y, z) \in E(T)$.

Suppose that $x$ is not a leaf in $T$. Vertex $z$ is the
parent of $x$ in $T'_r$, since every vertex is smaller than its parent in $\prec$.
Let $C_x$ denote the class of the partition containing $x$.
If $x$ and $z$ belong to different classes, then all vertices of $C_x$ are contained in $\dangling{T}{z}{x}$. Moreover, $x$ is the greatest vertex in $C_x$ in $\prec$, and thus $y$ cannot belong to $C_x$, contradicting the assumption.

On the contrary, suppose that $x$ and $z$ belong to the same class. The only vertices between $x$ and $z$ according to $\prec$ that belong to $C_x$ are leaves of $T$ adjacent to $z$. Thus, $y$ is adjacent to $z$. This completes the proof. 
\end{proof}

It was proved in \cite[Theorem 7]{B-D-thinness} that for a fixed value $m \in \NN$,
the thinness of a complete $m$-ary tree on $n$ vertices is
$\Theta(\log(n))$, and in \cite{Lucia-tesis}
that the thinness of a non-trivial tree is less than or equal to
its height. In contrast, computing the exact thinness of a complete $m$-ary tree 
was an open problem until now. As a consequence of
the \nameref{thm:characterization}, we have the following results.

\begin{theorem}[Thinness of complete $m$-ary trees]\label{thm:thin-m-ary}
    Let $m \geq 3$ and $T$ a complete $m$-ary tree with height $h$, then
    $\thin(T) = \ceil*{\frac{h+1}{2}}$.
\end{theorem}

\begin{proof} For a given $m \geq 3$ we proceed by induction on the height $h$ of an $m$-ary tree $T$.
    \begin{itemize}
    \item \emph{Base case} ($h \leq 1$):
    Tree $T$ is an interval graph, and thus has thinness 1. 
    The condition $\thin(T) = \ceil*{\frac{h+1}{2}}$ thus holds for both $h = 0$ and $h = 1$.

    \item \emph{Inductive step} ($h > 1$):
    Assume the property holds for all $m$-ary trees of height less than $h$.

    Let $x$ be the only vertex at level 1 of $T$; $v_1, v_2, \dots , v_m$ the level 2 vertices;
    and $t_{i,1}, t_{i,2}, \dots, t_{i,m}$ the vertices adjacent to \( v_i\) at level 3 for every $i\in\{1,\dots,m\}$. For every $j \in \{1,\dots,m\}$, tree 
    $\dangling{T}{v_i}{t_{i,j}}$ has height $h-2$ and is an $m$-ary tree.
    By the inductive hypothesis we have that
    $\thin(\dangling{T}{v_i}{t_{i,j}}) = \ceil*{\frac{h-1}{2}}$. 

    Since there are at least three vertices \( v_i\), each one with at least one dangling tree
    \( \dangling{T}{v_i}{t_{i,j}} \), $\abs*{\N{\ceil*{\frac{h-1}{2}}}{x}} \geq 3$. Due to the \nameref{thm:characterization}, we have that
    $\thin(T) \geq \ceil*{\frac{h-1}{2}}+1$. On the other hand, applying the \nameref{lem:path-layout}
    with the path containing only the vertex $x$ shows us that \( \thin(T) \leq \ceil*{\frac{h - 1}{2}} + 1 \). As \( \ceil*{\frac{h - 1}{2}} + 1 = \ceil*{\frac{h+1}{2}} \), we have that
    \( \ceil*{\frac{h+1}{2}} \leq \thin(T) \leq \ceil*{\frac{h+1}{2}} \), and so the property holds for $T$.
    \end{itemize}
\end{proof}

\begin{theorem}[Thinness of complete binary trees]\label{thm:thin-binary}
    Let $T$ be a complete binary tree with height $h$, then
    $\thin(T) = \ceil*{\frac{h+1}{3}}$.
\end{theorem}

\begin{proof} We proceed by induction on $h$.
    \begin{itemize}
    \item \emph{Base case} ($h \leq 2$):
    For $h \in \{0,1,2\}$, since \(T\) is an interval graph, $\thin(T) = 1 = \ceil*{\frac{h+1}{3}}$.

    \item \emph{Inductive step} ($h > 2$):
    Suppose that the property holds for all trees of height less than $h$. 
    
    We name the vertices of
    the first 4 levels of $T$ as described in \cref{fig:labels}, where $v_1$ is the root of $T$.

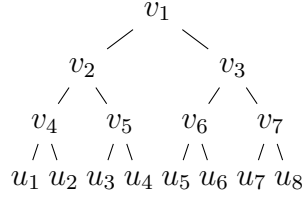
\begin{figure}[htbp]
     \centering
\begin{tikzpicture}[scale=0.5]
\tikzstyle{level 1}=[sibling distance=40mm] \tikzstyle{level
2}=[sibling distance=20mm] \tikzstyle{level 3}=[sibling
distance=10mm] \node{$v_1$}
    child{
        node{$v_2$}
            child{
                node{$v_4$}
                child{ node{$u_1$} }
                child{ node{$u_2$} }
            }
            child{
                node{$v_5$}
                child{ node{$u_3$} }
                child{ node{$u_4$} }
            }
    }
    child{
        node{$v_3$}
        child{
            node{$v_6$}
            child{ node{$u_5$} }
            child{ node{$u_6$} }
        }
        child{
            node{$v_7$}
            child{ node{$u_7$} }
            child{ node{$u_8$} }
        }
    };
\end{tikzpicture}
        \caption{Labels for some of the vertices of the first 4 levels of $T$.}
        \label{fig:labels}
\end{figure}

    Due to the inductive hypothesis we know that, for all $i \in \{4, 5, 6, 7\}$ and $j \in \NN$ such that $u_j$ is a neighbor of $v_i$, \[
    \thin(\dangling{T}{v_i}{u_j}) = \ceil*{\frac{h-3 + 1}{3}} = \ceil*{\frac{h-2}{3}}.
    \]
    As $\dangling{T}{v_6}{u_5}$ is an induced subgraph of $\dangling{T}{v_1}{v_3}$, by the \nameref{lemma:hereditary} we can also say that 
    \[
    \thin( \dangling{T}{v_1}{v_3} ) \geq  \thin( \dangling{T}{v_4}{u_1} ) = \ceil*{\frac{h-2}{3}}.
    \]
    
    These conditions imply that $\abs*{\N{\ceil*{\frac{h-2}{3}}}{v_2}} \geq 3$, so due to
    the \nameref{thm:characterization} we can say that \[
    \thin(T) \geq \ceil*{\frac{h-2}{3}} + 1 = \ceil*{\frac{h-2+3}{3}} = \ceil*{\frac{h+1}{3}}.
    \]
    
    We have proved that $\thin(T) \geq \ceil*{\frac{h+1}{3}}$. To prove that $\thin(T) \leq \ceil*{\frac{h+1}{3}}$ we can apply the \nameref{lem:path-layout} with \(P = (v_2, v_1, v_3)\) to deduce
    that \[
        \thin(T) \leq \ceil*{\frac{h-2}{3}}+1 =
    \ceil*{\frac{h-2+3}{3}} = \ceil*{\frac{h + 1}{3}}.
    \]
    \end{itemize}
\end{proof}

In~\cite{Lucia-tesis} it was shown that the thinness of a tree with diameter $d$ is at most $\frac{d}{2}$. Using the theorems above, we can improve this bound.

\begin{theorem}[Bound on the diameter]\label{diameter-bound}
    Let $T$ be a tree and $d$ its diameter, then $\thin(T) \leq \ceil*{\frac{d+1}{4}}$. Moreover, if the maximum degree of a vertex in $T$ is at most $3$, then $\thin(T) \leq  \ceil*{\frac{d+3}{6}}$.
\end{theorem}

\begin{proof} Let
$m$ be the maximum degree among all vertices of $T$. If $m \leq 2$, tree $T$ is an interval graph, and thus has thinness 1. The statement is satisfied in this case, as $\ceil*{\frac{d+3}{6}} \geq 1$. Suppose then that $m \geq 3$.

If $d$ is
even, consider the complete $m$-ary tree $T'$ with height
$h'=\frac{d}{2}$. Tree $T$ is isomorphic to an induced subgraph of $T'$. To see this, take $P$ to be a longest path in $T$, which has length $d + 1$. Root $T$ in the middle vertex $r$ of $P$ to obtain $T_r$. Tree $T_r$ has height $\frac{d}{2} = h'$, and each vertex of $T_r$ has at most $m$ children. Take $r$ to be the root of $T'$ in the isomorphism, and proceed recursively on the children of $r$ in $T_r$, assigning them to the children of the root in $T'$.

    By the \nameref{lemma:hereditary}, this implies that $\thin(T) \leq \thin(T')$.
    Now from the result proved in \cref{thm:thin-m-ary} we can say that
        $$ \thin(T') = \ceil*{\frac{h'+1}{2}} = \ceil*{\frac{ \frac{d}{2} +1}{2}} = \ceil*{\frac{d+2}{4}}$$
    By transitivity of $\leq$, if $d$ is even, $\thin(T) \leq \ceil*{\frac{d+2}{4}} = \ceil*{\frac{d+1}{4}}$.

    If $d$ is odd, consider two complete $(m-1)$-ary trees $T_1$ and $T_2$ with height $h''$, where $h'' =\frac{d-1}{2}$. Now consider 
    the tree $T''$ obtained by connecting the roots $T_1$ and $T_2$. Notice that every vertex that is not a leaf has degree $m$ in $T''$.
    The diameter of $T''$ is equal to $2h''+1$, which is equal to $d$. Let $P$ be a maximum length path in $T$, and $(u,v)$ the middle edge of that path. Every leaf $w$ of $T$ such that $u$ is not in the path between $w$ and $v$ is at distance at most $h''$ from $v$,
    because otherwise there would be a path of length greater than $d$ in $T$. Symmetrically, every leaf $w$ of $T$ such that $v$ is not in the path between $w$ and $u$ is at distance at most $h''$ from $u$. Dangling trees $\dangling{T}{u}{v}$ and $\dangling{T}{v}{u}$ are then isomorphic to some induced subgraph of $T_1$ and $T_2$, respectively. This can be shown by defining $v$ and $u$ as the root of $T_1$ and $T_2$ respectively. As $T$ is exactly the result of connecting $\dangling{T}{u}{v}$ and $\dangling{T}{v}{u}$ by an edge on their respective roots, $T$ is isomorphic to an induced subgraph of $T''$.

    This implies that $\thin(T) \leq \thin(T'')$.
    Now from the \nameref{lem:path-layout} applied to the path $(u, v)$, and the result proved in \cref{thm:thin-m-ary}, we can say that
        $$ \thin(T'') \leq \ceil*{\frac{h''-1}{2}} + 1= \ceil*{\frac{ \frac{d-1}{2} -1}{2}} + 1 = \ceil*{\frac{d-3}{4}} + 1 = \ceil*{\frac{d+1}{4}}.$$
    By transitivity, if $d$ is odd, $\thin(T) \leq \ceil*{\frac{d+1}{4}}$. Combining both cases we obtain that $\thin(T) \leq \ceil*{\frac{d+1}{4}}$ for every tree $T$ with diameter $d \in \mathbb{N}$.

    For $m = 3$ and $d$ odd, we can use the same construction as earlier, and apply \cref{thm:thin-binary} instead of \cref{thm:thin-m-ary} to obtain $$
    \thin(T) \leq \ceil*{\frac{\frac{d-1}{2}-1}{3}} + 1= \ceil*{\frac{d+3}{6}}.$$ 
    
    For $d$ even, we make a similar construction by joining the roots of three
complete binary trees of height $\frac{d}{2}$ to a new vertex $x$. Applying the \nameref{lem:path-layout} to the path that only contains $x$ results in $\thin(T) \leq \ceil*{\frac{d+4}{6}} = \ceil*{\frac{d+3}{6}}$. Again, we can combine the even and
odd case and say that for every tree $T$ with diameter $d$ and
maximum degree at most $3$, $\thin(T) \leq \ceil{\frac{d+3}{6}}$.
\end{proof}

\section{Computing the thinness of a tree}

In this section, we present an algorithm that will compute the thinness of a tree, along with an optimal consistent solution. First, we define the following.

\begin{definition}[complete rooted subtree]
    We define the \emph{complete rooted subtree} $\rooted{T_r}{x}$ of $T_r$ as the subtree of $T_r$ rooted at $x$ induced by $x$ and the descendants of $x$.
\end{definition}

\begin{definition}[child and grandchild subtree]
    Let $x$ be a vertex of a rooted tree $T_r$. For every child $v$ of $x$, we say that $\rooted{T_r}{v}$ is a \emph{child subtree} of $x$ in $T_r$. For every grandchild $w$ of $x$, we say that $\rooted{T_r}{w}$ is a \emph{grandchild subtree} of $x$ in $T_r$.
\end{definition}

The algorithm will start by defining an arbitrary vertex $r$ as the root, and then proceed bottom-up, computing the thinness of a complete rooted subtree using the thinness of its child and grandchild subtrees. When the algorithm finishes, the thinness of the tree will be stored in the root $r$. 

The algorithm will rely heavily on the location of vertices we call \emph{critical}.

\begin{definition}[child $k$-neighbor]
    Let $T_r$ be a tree rooted on $r$. A vertex $v \in V(T_r)$ is a \emph{child $k$-neighbor} of $x$ if $v$ is both a $k$-neighbor and a child of $x$.
\end{definition}

\begin{definition}[critical vertex]
    Let $T_r$ be a rooted tree with thinness $k$. We call a vertex $x$ in $T_r$ \emph{critical} if it has exactly two child $k$-neighbors.
\end{definition}

Critical vertices are important because, if a child subtree of a vertex $r$ has thinness $k$ and has a critical vertex $v$, then $v$ is a candidate to being $k$-saturated in $T_r$. This and other properties of critical vertices are formalized in the following propositions.

\begin{observation}
    \label{obs:parent-of-critical-vertex-is-not-k-neighbor}
    Let $T_r$ be a tree rooted on $r$ with thinness $k$. The parent of a critical vertex $x \in V(T_r)$ cannot be a $k$-neighbor of $x$.
\end{observation}
\begin{proof}
    The only neighbors of $x$ are its children and its parent. As $x$ is critical, exactly two of the children of $x$ are $k$-neighbors of $x$. Thus, if the parent of $x$ is also a $k$-neighbor of $x$, then $\abs*{\N{k}{x}} \geq 3$. Therefore, by the \nameref{thm:characterization}, $T_r$ has thinness at least $k+1$, which is a contradiction. 
\end{proof}

\begin{lemma}
\label{lemma:one-critical-vertex}
    Every rooted tree with thinness $k$ has at most one critical vertex.
\end{lemma}

\begin{proof}
We will prove the statement by contradiction. Let $T_r$ be a tree rooted on $r$ with thinness $k$.
Let $x$ and $y$ be two different critical vertices in $T_r$.

Take the simple path $\simplePath(x, y)$. Note that $\simplePath(x, y)$ contains either the parent of $x$, the parent of $y$, or both. Suppose, without loss of generality, that $\simplePath(x, y)$ contains the parent of $x$, which means that the two child $k$-neighbors of $x$ do not belong to $\simplePath(x, y)$. Furthermore, at least one of the child $k$-neighbors of $y$ is not contained in $\simplePath(x, y)$. Therefore, path $\simplePath(x, y)$ has at least 3 $k$-neighbors, and by \cref{coro:characterization-by-paths}, $T_r$ has thinness at least $k+1$, which is a contradiction.
\end{proof}

\Cref{lemma:one-critical-vertex} has strong implications for our algorithm. For example, if for some $k$ we have that at least two child subtrees with thinness $k$ have critical vertices, then we know that the thinness of the tree is at least $k+1$. The following lemma will constrain the possible thinness values of the tree even further.

\begin{lemma}
    \label{lemma:rooted-tree-bound}
    For every non-trivial tree $T_r$ rooted on $r$, we have $k \leq \thin(T_r) \leq k+1$, where $k = \max_{v \in \child(r)} \{ \thin(\rooted{T_r}{v}) \}$, meaning, $k$ is the maximum thinness of a child subtree of $r$. 
\end{lemma}
\begin{proof}
    At least one child subtree of $r$ has thinness $k$, so by the \nameref{lemma:hereditary}, $k \leq \thin(T_r)$. 
    
    On the other hand, take the path $P$ that contains only $r$. The connected components of $T_r \setminus N[P]$ are subtrees of the child subtrees of $r$. By the \nameref{lemma:hereditary}, the thinness of each of these subtrees is at most $k$. Thus, by the \nameref{lem:path-layout}, $\thin(T_r) \leq k+1$.
\end{proof}

We are now ready to show the main structure of the procedure that will be executed in each vertex. We present it as the decision tree in \vref{fig:decision-tree}.

\tikzset{rect/.style={draw, rectangle, font=\footnotesize}, align=center}
\tikzset{condition/.style={draw, chamfered rectangle, font=\footnotesize, align=center, inner sep=1pt}}
\tikzset{line/.style={draw, -latex}}
\newcommand*{\nodetext}[3][4cm]{\underline{\emph{#2}}\\\parbox[t]{#1}{#3}}
\begin{figure}
\centering
\begin{tikzpicture}[auto]
    \node[rect] (start) {Start};
    \node[condition, below=0.7cm of start] (cond1) {
        \underline{\emph{Condition 1}}\\
        Is $r$ a leaf?
    };
    \node[condition, below=0.7cm of cond1] (cond2) {
        \nodetext[2.5cm]{Condition 2}{
        Is $\abs*{\N[T_r]{k}{r}} \geq 3$?
        }
    };
    \node[condition, below=0.7cm of cond2] (cond3) {
        \nodetext{Condition 3}{
        Is there a child subtree of $r$
        with thinness $k$ that contains
        a critical vertex?
        }
    };
    
    \node[condition, below=0.7cm of cond3] (cond4) {
        \nodetext[3.6cm]{Condition 4}{
        Is there more than one child subtree of $r$ with thinness $k$?
        }
    };
    \node[rect, below=0.7cm of cond4] (assign-v) {
        \parbox[t]{6cm}{
        Let $v$ be the child of $r$ such that $\rooted{T_r}{v}$ has
        thinness $k$, and $\rooted{T_r}{v}$ has a single critical vertex $x$ (there is only one critical vertex by \cref{lemma:one-critical-vertex}).
        }
    };
    \node[condition, below=0.7 of assign-v] (cond5) {
        \nodetext[3cm]{Condition 5}{
        Is $v$ the critical vertex of $\rooted{T_r}{v}$?
        }
    };
    \node[condition, below=0.7 of cond5] (cond6) {
        \nodetext{Condition 6}{
        Is $\thin(T_r \setminus \rooted{T_r}{w}) \geq k$ for $w$ parent of $x$?
        }
    };

    \node[rect, right=0.7cm of cond1] (case1) {
        \underline{\emph{Case 1}}\\
        $\thin(T_r) = 1$
    };
    \node[rect, right=0.7cm of cond2] (case2) {
        \underline{\emph{Case 2}}\\
        $\thin(T_r) = k + 1$
    };
    \node[rect, right=0.7cm of cond3] (case3) {
        \underline{\emph{Case 3}}\\
        $\thin(T_r) = k$
    };
    \node[rect, right=0.7cm of cond4] (case4) {
        \underline{\emph{Case 4}}\\
        $\thin(T_r) = k + 1$
    };
    \node[rect, right=0.7cm of cond5] (case5) {
        \underline{\emph{Case 5}}\\
        $\thin(T_r) = k$
    };
    \node[rect, right=0.7cm of cond6] (case6) {
        \underline{\emph{Case 6}}\\
        $\thin(T_r) = k + 1$
    };
    \node[rect, below=0.7cm of cond6] (case7) {
        \underline{\emph{Case 7}}\\
        $\thin(T_r) = k$
    };

    \path[line] (start) -- (cond1);
    \path[line] (cond1) -- (case1) node [midway, font=\footnotesize] {Yes};
    \path[line] (cond1) -- (cond2) node [midway, font=\footnotesize] {No};
    \path[line] (cond2) -- (cond3) node [midway, font=\footnotesize] {No};
    \path[line] (cond2) -- (case2) node [midway, font=\footnotesize] {Yes};
    \path[line] (cond3) -- (case3) node [midway, font=\footnotesize] {No};
    \path[line] (cond3) -- (cond4) node [midway, font=\footnotesize] {Yes};
    \path[line] (cond4) -- (case4) node [midway, font=\footnotesize] {Yes};
    \path[draw] (cond4) -- (assign-v) node [midway, font=\footnotesize] {No};
    \path[line] (assign-v) -- (cond5);
    \path[line] (cond5) -- (case5) node [midway, font=\footnotesize] {Yes};
    \path[line] (cond5) -- (cond6) node [midway, font=\footnotesize] {No};
    \path[line] (cond6) -- (case6) node [midway, font=\footnotesize] {Yes};
    \path[line] (cond6) -- (case7) node [midway, font=\footnotesize] {No};
\end{tikzpicture}
\caption{A decision tree to compute the thinness of a rooted tree $T_r$. Here, $k$ is the maximum thinness of a child subtree of $r$; in other words, $k = \max_{v \in \child(r)} \{ \thin(\rooted{T_r}{v}) \}$.}
\label{fig:decision-tree}
\end{figure}
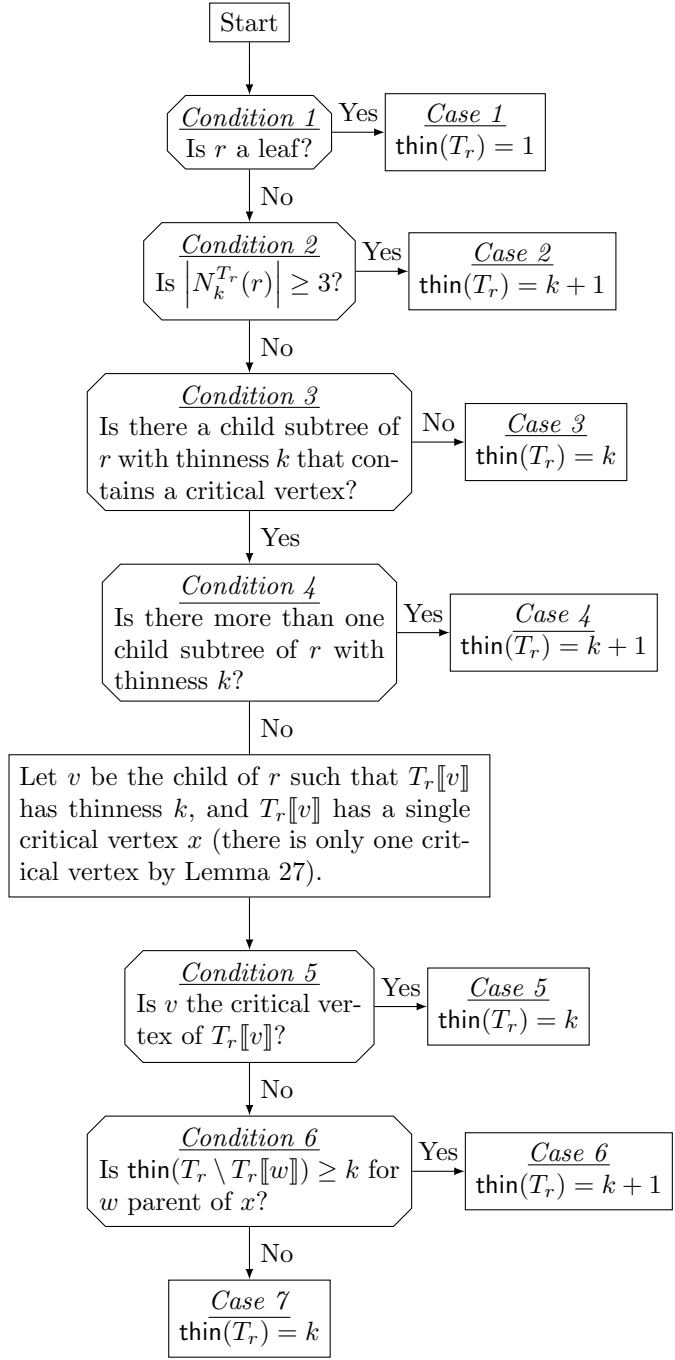

\begin{lemma}
\label{lemma:decision-tree}
    The decision tree in \cref{fig:decision-tree} correctly computes the thinness of a tree $T_r$ rooted on $r$.
\end{lemma}
\begin{proof}
    Let $k$ be the maximum thinness of a child subtree of $r$. We go case by case, proving that the thinness of $T_r$ is the one stated.
    \begin{itemize}
        \item \emph{Case 1}: This case is only reached when $r$ is a leaf. Clearly, there are no $1$-saturated vertices, and thus by the \nameref{thm:characterization} $\thin(T_r) = 1$.
    \end{itemize}
    In the following cases, $T_r$ must be a non-trivial tree. Therefore, by \cref{lemma:rooted-tree-bound} and the \nameref{thm:characterization}, the presence or absence of a $k$-saturated vertex in $T_r$ is sufficient to determine the thinness of $T_r$.
    \begin{itemize}
        \item \emph{Case 2}: This case is reached only if $\abs*{\N[T_r]{k}{r}} \geq 3$, which is the definition of a $k$-saturated vertex.
        \item \emph{Case 3}: In this case $r$ is not $k$-saturated, as $\abs*{\N[T_r]{k}{r}} \leq 2$. By the \nameref{thm:characterization}, no vertex $v$ contained in a child subtree of $r$ can have more than two child $k$-neighbors. Moreover, there is no child subtree of $r$ with thinness $k$ that contains a critical vertex, and thus $v$ has less than two child $k$-neighbors. Therefore, $v$ is not $k$-saturated. Hence, there are no $k$-saturated vertices in $T_r$.
        \item \emph{Case 4}: In this case there is at least one child subtree of $r$ with thinness $k$ that has a critical vertex, and another child subtree of $r$ with thinness $k$. Let $v$ be a child of $r$ such that $\rooted{T_r}{v}$ has thinness $k$, and $\rooted{T_r}{v}$ has a critical vertex $x$. Let $u$ be a child of $r$ different from $v$ such that $\rooted{T_r}{u}$ has thinness $k$. Note that $r$ is a $k$-neighbor of $v$, as $\rooted{T_r}{u} = \dangling{T_r}{r}{u}$. The path $\simplePath(x, v)$ thus has at least three $k$-neighbors: the two child $k$-neighbors of $x$, and the $k$-neighbor $r$ of $v$. Therefore, by \cref{coro:characterization-by-paths}, $T_r$ has thinness $k+1$.
    \end{itemize}
    The following cases are only reached if there is only one child $v$ of $r$ such that $\rooted{T_r}{v}$ has thinness $k$, and $\rooted{T_r}{v}$ has a critical vertex $x$. Clearly, $r$ cannot be $k$-saturated, and neither can any vertex with less than two child $k$-neighbors. Therefore, the only potential $k$-saturated vertex is $x$. In fact, we only have to check if the parent of $x$ is its $k$-neighbor to decide if it is $k$-saturated.
    \begin{itemize}
        \item \emph{Case 5}: This case is reached when $x = v$. The parent of $v$ is $r$. The only dangling trees from $r$ are the child subtrees of $r$, and the only child subtree of $r$ with thinness $k$ is $\rooted{T_r}{v}$. This leaves us with the conclusion that $r$ is not a $k$-neighbor of $v$, and thus there is no $k$-saturated vertex in $T_r$.
        \item \emph{Case 6 and 7}: Checking if $x$ is $k$-saturated is equivalent to checking if there is a dangling subtree from its parent $w$ different from $\dangling{T_r}{w}{x}$ that has thinness $k$. 
        
        First, we show that for every child $y$ of $w$ different from $x$ we have that $\thin(\dangling{T_r}{w}{y}) < k$. Suppose that $\thin(\dangling{T_r}{w}{y}) \geq k$. This implies that $w$ is a $k$-neighbor of $x$ in $\rooted{T_r}{v}$, because $\dangling{T_r}{w}{y}$ is an induced subgraph of $\rooted{T_r}{v}$. Thus, $x$ is $k$-saturated in $\rooted{T_r}{v}$, and $\rooted{T_r}{v}$ has thinness greater than $k$ by the \nameref{thm:characterization}. This is a contradiction.

        Now, let $u$ be the parent of $w$, which exists because $x$ is not $v$ in these cases. From the previous statement we deduce that $w$ is a $k$-neighbor of $x$ if and only if the tree $\dangling{T_r}{w}{u}$ has thinness at least $k$. This tree is exactly $T_r \setminus \rooted{T_r}{w}$. Therefore, $x$ is $k$-saturated if and only if $\thin(T_r \setminus \rooted{T_r}{w}) \geq k$, and Cases 6 and 7 compute the thinness of $T_r$ accordingly.
    \end{itemize}
\end{proof}

All conditions in \cref{lemma:decision-tree} are easy to compute given the thinness and critical vertices of the child subtrees of $r$, except for Condition 6. This condition requires to know the thinness of a subtree of $T_r$ that is not a complete rooted subtree of $T_r$, namely, $T_r \setminus \rooted{T_r}{w}$. Thus, a naive bottom-up algorithm, meaning, one that only computes the thinness and critical vertex of the complete rooted subtree that is being processed, will not be sufficient for our purposes.

Suppose that we tried computing the thinness of $T_r \setminus \rooted{T_r}{w}$ by calling this algorithm recursively. We would need to repeat the whole bottom-up procedure for subtree $T_r \setminus \rooted{T_r}{w}$, which would be inefficient. Instead, notice that the only information needed by Conditions 1 through 5 to compute the thinness of $T_r \setminus \rooted{T_r}{w}$ is the thinness of its child and grandchild subtrees. Moreover, the only child and grandchild subtrees that are modified when removing $\rooted{T_r}{w}$ are the ones that contain $\rooted{T_r}{w}$. Thus, we only need to compute the thinness of that specific child and grandchild subtrees, as the algorithm would have already computed the thinness of the other ones. Still, this is not enough, as it involves restarting the algorithm on the child and grandchild subtrees that contained $\rooted{T_r}{w}$ each time Condition 6 is reached.

Luckily, there is still one optimization to be made, led by the following observation.

\begin{observation}
    \label{obs:critical-vertex-of-parent-is-same-as-descendant}
    Let $T_r$ be a tree rooted on a vertex $r$ with a critical vertex $x$. If $v$ is an ancestor of $x$, then $x$ is also a critical vertex of $\rooted{T_r}{v}$.
\end{observation}
\begin{proof}
    Let $k$ be the thinness of $T_r$. As $x$ is a critical vertex of $T_r$, it has exactly two child $k$-neighbors. Hence, by the \nameref{lemma:hereditary}, $\rooted{T_r}{x}$ has thinness $k$, and so does $\rooted{T_r}{v}$. Therefore, $x$ is also a critical vertex of $\rooted{T_r}{v}$.
\end{proof}

Suppose that $v$ is a child of $r$ such that $w \in V(\rooted{T_r}{v})$. Then $v$ is an ancestor of $x$, and by \cref{obs:critical-vertex-of-parent-is-same-as-descendant}, $x$ is the critical vertex of $\rooted{T_r}{v}$. This means that Condition 6 was reached when processing $\rooted{T_r}{v}$. If $x$ is a child of $v$, then $v = w$, and $\rooted{T_r}{v} \setminus \rooted{T_r}{w}$ is simply empty. Otherwise, the thinness of $\rooted{T_r}{v} \setminus \rooted{T_r}{w}$ was already required when processing $\rooted{T_r}{v}$, and thus we can reuse this value.

The solution is then simple: after processing $\rooted{T_r}{v}$, we return not only its thinness, but also the thinness of $\rooted{T_r}{v} \setminus \rooted{T_r}{w}$. It could be the case that we also reach Condition 6 when processing $\rooted{T_r}{v} \setminus \rooted{T_r}{w}$. If this happens, there will be another critical vertex in $\rooted{T_r}{v} \setminus \rooted{T_r}{w}$ with parent $w'$, and we will need to compute the thinness of $(\rooted{T_r}{v} \setminus \rooted{T_r}{w}) \setminus \rooted{T_r}{w'}$. In this case, we also return the thinness of $(\rooted{T_r}{v} \setminus \rooted{T_r}{w}) \setminus \rooted{T_r}{w'}$ after processing $\rooted{T_r}{v}$, and so on.

This hypothetical algorithm needs to return more than just the thinness value for this to work. It must return a list of thinness values. Also, we need to know where the critical vertex is located in each of the subtrees in this sequence, and thus need to return a list of critical vertices along with the list of thinness values. We define these two lists as follows.

\begin{definition}[critical vertex list]
    \label{def:critical-vertex-list}
    Let $T_r$ be a tree rooted on $r$. The \emph{critical vertex list of $T_r$}, denoted $\critList(T_r)$, is a list of vertices of $T_r$ or a different arbitrary value \nil\ defined recursively as follows:
    \begin{itemize}
        \item If $T_r$ has no critical vertex, then $\critList(T_r) = (\nil)$.
        \item Otherwise, let $x$ be the critical vertex in $T_r$. If $x$ is $r$ or a child of $r$, then $\critList(T_r) = (x)$.
        \item Otherwise, if $w$ is the parent of $x$, then $\critList(T_r) = x \oplus \critList(T_r \setminus \rooted{T_r}{w})$.
    \end{itemize}
\end{definition}

We can now give a name to the subtrees mentioned earlier.

Let $\ell = \abs{\critList(T_r)}$, and let $w_1,\dots,w_{\ell - 1}$ be the parents of the first $\ell - 1$ vertices in $\critList(T_r)$.
For every $j \in \{1,\dots,\ell\}$, we will use $\subtreeWithoutCriticals{T_r}{j}$ to denote the tree $T_r \setminus \rooted{T_r}{w_1} \setminus \dots \setminus \rooted{T_r}{w_{j-1}}$. Note that $\subtreeWithoutCriticals{T_r}{1} = T_r$.

From the \defref{def:critical-vertex-list} we can derive the following observation. 
\begin{observation}
    \label{obs:suffix-of-crit-list-is-crit-list-of-Trj}
    Let $T_r$ be a tree rooted on $r$, and $\ell = \abs{\critList(T_r)}$.
    For every $i \in \{1,\dots,\ell\}$, the suffix starting at position $i$ of $\critList(T_r)$ is equal to $\critList(\subtreeWithoutCriticals{T_r}{i})$.
\end{observation}
\begin{proof}
    We will prove this by induction on $i$.
    \begin{itemize}
        \item \emph{Base case} ($i = 1$): In this case, $\suffix{\critList(T_r)}{i} = \critList(T_r)$, and thus the statement holds, as $\subtreeWithoutCriticals{T_r}{1} = T_r$.
        \item \emph{Inductive step} ($i > 1$): Assume the statement holds for $i - 1$, meaning, \[\suffix{\critList(T_r)}{i-1} = \critList(\subtreeWithoutCriticals{T_r}{i-1}).\] We will show that $\suffix{\critList(T_r)}{i} = \critList(\subtreeWithoutCriticals{T_r}{i})$. Note that this is the same as proving that $\suffix{\critList(\subtreeWithoutCriticals{T_r}{i-1})}{2} = \critList(\subtreeWithoutCriticals{T_r}{i})$.
        
        As $i \leq \ell$, the length of $\critList(\subtreeWithoutCriticals{T_r}{i-1})$ is at least 2. By the \defref{def:critical-vertex-list}, $\subtreeWithoutCriticals{T_r}{i-1}$ has a critical vertex $v$ such that \[\critList(\subtreeWithoutCriticals{T_r}{i-1}) = v \oplus \critList(\subtreeWithoutCriticals{T_r}{i}).\] Thus, $\suffix{\critList(\subtreeWithoutCriticals{T_r}{i-1})}{2} = \critList(\subtreeWithoutCriticals{T_r}{i})$, and the statement holds.
    \end{itemize}
\end{proof}

With this we see that $\critList(T_r)$ contains each of the critical vertices of the subtrees $\subtreeWithoutCriticals{T_r}{1}, \subtreeWithoutCriticals{T_r}{2},$ etc.

\begin{definition}[thinness list]
    \label{def:thinness-list}
    Let $T_r$ be a tree rooted on $r$ with thinness $k$. Let $\ell$ be the length of $\critList(T_r)$. The \emph{thinness list of $T_r$}, denoted $\thinList(T_r)$, is a list of $\ell$ natural numbers where $\thinList(T_r)_i$ is the thinness of $\subtreeWithoutCriticals{T_r}{i}$ for every $i \in \{1,\dots,\ell\}$.
\end{definition}

Note that the first element of $\thinList(T_r)$ is always the thinness of $T_r$.

\begin{observation}
    \label{obs:suffix-of-thinness-list-is-thinness-list-of-Tr2}
    Let $T_r$ be a rooted tree. If $\thinList(T_r)$ has more than one element, then $\suffix{\thinList(T_r)}{2} = \thinList(\subtreeWithoutCriticals{T_r}{2})$.
\end{observation}
\begin{proof}
    The definition of $\subtreeWithoutCriticals{T_r}{2}$ implies that $\subtreeWithoutCriticals{\left(\subtreeWithoutCriticals{T_r}{2}\right)}{i} = \subtreeWithoutCriticals{T_r}{i+1}$ for every $i \in \{1,\dots,\ell-1\}$. From this we arrive directly to the statement.
\end{proof}

Now, let us revisit our previous idea. Suppose that we are processing a rooted tree $T_r$ and Condition 6 is reached. Additionally, suppose that the critical vertex $x$ in the child subtree $\rooted{T_r}{v}$ is not a child of $v$, but instead has a different parent, denoted $w$. In our hypothetical algorithm, the thinness and the critical vertex of $\rooted{T_r}{v} \setminus \rooted{T_r}{w}$ were already computed when processing $\rooted{T_r}{v}$, and so we can store this information to reuse it when processing Condition 6. In particular, the information will be stored as the second element of the lists we just defined. We formalize this in the following lemma.

\begin{lemma}
    \label{lemma:Trv2-is-child-subtree-of-Tr-minus-Trw}
    Let $T_r$ be a tree rooted on a vertex $r$ such that $\abs{\critList(T_r)} \geq 2$. Let $v$ be a child of $r$ such that $\rooted{T_r}{v}$ has a critical vertex that is neither $v$ nor a child of $v$. If $w$ is the parent of that critical vertex, then \[
        (T_r \setminus \rooted{T_r}{w}) \cap \rooted{T_r}{v} = \subtreeWithoutCriticals{\rooted{T_r}{v}}{2}.
    \]
    In particular, $\subtreeWithoutCriticals{\rooted{T_r}{v}}{2}$ is a child subtree of $r$ in $T_r \setminus \rooted{T_r}{w}$.
\end{lemma}
\begin{proof}
    By the \defref{def:critical-vertex-list}, the critical vertex list of $\rooted{T_r}{v}$ has at least two elements, which means that $\subtreeWithoutCriticals{\rooted{T_r}{v}}{2}$ is well-defined. Given that $\rooted{T_r}{v}$ is a child subtree of $T_r$, we have that $T_r \cap \rooted{T_r}{v} = \rooted{T_r}{v}$. We thus have to prove only that \[\rooted{T_r}{v} \setminus \rooted{T_r}{w} = \subtreeWithoutCriticals{\rooted{T_r}{v}}{2}.\] In particular, we have to prove that the first element of the critical vertex list of $\rooted{T_r}{v}$ has $w$ as its parent. This element is the critical vertex of $\rooted{T_r}{v}$. As we selected $w$ exactly to be the parent of the critical vertex in $\rooted{T_r}{v}$, the statement holds.
\end{proof}

In our algorithm, we will potentially make a recursive call for every element of $\thinList(T_r)$ when processing $T_r$. Thus, it is important to set an upper bound on the number of elements of $\thinList(T_r)$. We do this with the following two lemmas.

\begin{lemma}
    \label{lemma:thinness-of-Tri-is-bigger-than-thinness-of-Tri+1}
    Let $T_r$ be a rooted tree, and let $\ell = \abs{\critList(T_r)}$. For every $i \in \{1,\dots,\ell-1\}$, we have that $\thin(\subtreeWithoutCriticals{T_r}{i}) > \thin(\subtreeWithoutCriticals{T_r}{i+1})$.
\end{lemma}
\begin{proof}
    By definition, $\subtreeWithoutCriticals{T_r}{i}$ has a critical vertex $v$ with parent $w$. Let $k = \thin(\subtreeWithoutCriticals{T_r}{i})$. By \cref{obs:parent-of-critical-vertex-is-not-k-neighbor}, $w$ cannot be a $k$-neighbor of $v$. In particular, $\subtreeWithoutCriticals{T_r}{i} \setminus \rooted{T_r}{w}$ has thinness lower than $k$. This is exactly $\subtreeWithoutCriticals{T_r}{i+1}$, and so the statement holds.
\end{proof}

\begin{lemma}
    \label{lemma:thinness-list-has-log-n-elements}
    The thinness list of a rooted tree on $n$ vertices has $\bigO(\log(n))$ elements.
\end{lemma}
\begin{proof}
    Let $T_r$ be a rooted tree. By \cref{log-bound}, the thinness of $T_r$, which is the first element of its thinness list, has order of magnitude $\bigO(\log(n))$. By the \defref{def:thinness-list} and \cref{lemma:thinness-of-Tri-is-bigger-than-thinness-of-Tri+1}, the thinness list of $T_r$ is a strictly decreasing list. The thinness of a subtree cannot be lower than 1, and therefore the thinness list of $T_r$ has at most $\bigO(\log(n))$ elements.
\end{proof}

We are now ready to describe the algorithm in detail. First, we give the procedure \textproc{ComputeThinnessAndCritLists} that will compute the thinness list and the critical vertex list of a rooted tree given the thinness lists and the critical vertex lists of the child and grandchild subtrees of the root.

\begin{lemma}
    \label{lemma:compute-list-and-type}
    Let $T_r$ be a tree rooted on a vertex $r$. Let $n = \abs{V(T_r)}$. Given the critical vertex lists and the thinness lists of the child and grandchild subtrees of $r$, the critical vertex list and the thinness list of $T_r$ can be computed in $\bigO((\abs{\child(r)} + \abs{\gchild(r)}) \cdot \log(n))$ time.
\end{lemma}
\begin{proof}
    Let $T^\star_r$ be a supertree of $T_r$ rooted on the same vertex $r$. We define the recursive procedure \[
        \textproc{ComputeThinnessAndCritLists}(T^\star_r, \childInfo, \gchildInfo)
    \]
    that will compute the critical vertex list and the thinness list of $T_r$ given the root $r$ and the information on the child and grandchild subtrees of $r$ in $T_r$ contained in lists $\childInfo$ and $\gchildInfo$. The information on a complete subtree rooted on $v$ is a triple of the form \[\angles{v, \critList(\rooted{T_r}{v}), \thinList(\rooted{T_r}{v})}.\] For each child $v$ of $r$, the list $\childInfo$ contains the information on $\rooted{T_r}{v}$, and for each grandchild $w$ of $r$, the list $\gchildInfo$ contains the information on $\rooted{T_r}{w}$.
    
    We only require that $T^\star_r$ is a supertree of $T_r$ because we will use this procedure recursively on subtrees of $T_r$, and to maintain the required time complexity we need to avoid creating new subtrees to pass as arguments to these recursive calls. The data structure used for $T^\star_r$ must allow determining in constant time both the parent and the children of a vertex.

    The procedure \textproc{ComputeThinnessAndCritLists} follows the decision tree shown in \vref{fig:decision-tree}. When it reaches a case, it computes the critical vertex list and the thinness list of $T_r$. We now show how to check each condition in the decision tree and compute the critical vertex list and the thinness list in the time complexity required by the statement.

    We start by showing that the conditions in the tree can each be asserted in $\bigO((\abs{\child(r)} + \abs{\gchild(r)}) \cdot \log(n))$ time. The only condition that requires a recursive call is Condition 6, and thus trees that reach conditions 1 through 5 will be base cases in our procedure.
    \begin{itemize}
        \item \emph{Condition 1: Is $r$ a leaf?} This only happens when $r$ has no children, which can be asserted by checking if $\childInfo$ is empty. This can be done in constant time.
        \item \emph{Condition 2: Is $\abs*{\N[T_r]{k}{r}} \geq 3$?} First, recall that $k$ is the maximum thinness of a child subtree of $r$. We can compute $k$ by iterating over all the elements of $\childInfo$ and taking the maximum over the first element of the thinness lists. This can be done in $\bigO(\abs{\child(r)})$ time.
        
        For each grandchild $g$ of $r$ in $\gchildInfo$, we can compute which vertex is the parent of $g$ by consulting the data structure used for $T^\star_r$. As $T^\star_r$ is a supertree of $T_r$, the parent of $g$ in $T^\star_r$ is the same as in $T_r$.
        
        We can then go through the $\gchildInfo$ list and count the number of children $v$ of $r$ that have at least one child $g$ such that $\rooted{T_r}{g}$ has thinness $k$. If there are at least three, then the condition is satisfied, and it is not otherwise. This can be done in $\bigO(\abs{\gchild(r)})$ time.
        
        \item \emph{Condition 3: Is there a child subtree of $r$ with thinness $k$ that contains a critical vertex?} We can iterate over all the entries in $\childInfo$ to find the children $v$ of $r$ such that $\thin(\rooted{T_r}{v}) = k$ and that $\critList(\rooted{T_r}{v}) \neq (\nil)$. This can be done in $\bigO(\abs{\child(r)})$ time.

        \item \emph{Condition 4: Is there more than one child subtree of $r$ with thinness $k$?} This can be similarly asserted by iterating over all the elements of $\childInfo$ and checking if there are at least two child subtrees of $r$ with thinness $k$. This can be done in $\bigO(\abs{\child(r)})$ time.
        
        \item \emph{Condition 5: Is $v$ the critical vertex of $\rooted{T_r}{v}$?} 
        We first iterate over all the elements of $\childInfo$ to find the child $v$ of $r$ such that $\rooted{T_r}{v}$ has thinness $k$ and $\rooted{T_r}{v}$ has a critical vertex.
        Then, we check if the first element of $\critList(\rooted{T_r}{v})$ is equal to $v$.
        This can be done in $\bigO(\abs{\child(r)})$ time.
        
        \item \emph{Condition 6: Is $\thin(T_r \setminus \rooted{T_r}{w}) \geq k$ for $w$ parent of $x$?} We need to compute the thinness of a subtree of $T_r$ that is not a complete rooted subtree, namely, $T_r \setminus \rooted{T_r}{w}$. Let $T'_r$ denote this subtree. To solve this, we will make a recursive call to \textproc{ComputeThinnessAndCritLists} with tree $T^\star_r$ and the information on the child and grandchild subtrees of $r$ in $T'_r$. Note that $T^\star_r$ is a supertree of $T'_r$, as $T'_r$ is a subtree of $T_r$, which itself is a subtree of $T^\star_r$.
        
        For every child $v'$ of $r$ in $T_r$, we know that $\rooted{T_r}{v'} \cap T'_r$ is a child subtree of $r$ in $T'_r$, if $\rooted{T_r}{v'} \cap T'_r$ is not empty. Thus, we have that $\rooted{T'_r}{v'} = \rooted{T_r}{v'} \setminus \rooted{T_r}{w}$.
        
        Recall that $v$ is the child of $r$ such that $\rooted{T_r}{v}$ contains $x$. If $v'$ is different from $v$, then $\rooted{T_r}{v'}$ does not contain any vertex of $\rooted{T_r}{w}$. Therefore, $\rooted{T'_r}{v'} = \rooted{T_r}{v'}$. We can thus maintain the same information for $\rooted{T'_r}{v'}$ on the recursive call without modification.
        
        We know by Condition 5 that $v$ is not the critical vertex of $\rooted{T_r}{v}$. If a child of $v$ is the critical vertex $x$, then the parent $w$ of $x$ is equal to $v$, and then $\rooted{T'_r}{v}$ is empty. We can therefore eliminate the information on $\rooted{T_r}{v}$ from $\childInfo$ and the information on the child subtrees of $v$ from $\gchildInfo$ to obtain the full child and grandchild information for $T'_r$. We will store this data in new variables so as to restore it to $\childInfo$ and  $\gchildInfo$ after the recursive call. It is important to just modify $\childInfo$ and $\gchildInfo$ instead of making a copy of those variables in the recursive call to maintain the runtime of the algorithm.

        If, to the contrary, the critical vertex $x$ of $\rooted{T_r}{v}$ is not a child of $v$, then by \cref{lemma:Trv2-is-child-subtree-of-Tr-minus-Trw} we know that $\subtreeWithoutCriticals{\rooted{T_r}{v}}{2} = \rooted{T'_r}{v}$. Furthermore, by the \defref{def:critical-vertex-list}, the critical vertex list of $\subtreeWithoutCriticals{\rooted{T_r}{v}}{2}$ is equal to $\suffix{\critList(\rooted{T_r}{v})}{2}$. Therefore, we only need to remove the first element of $\critList(\rooted{T_r}{v})$ to obtain the critical vertex list of $\rooted{T'_r}{v}$. Similarly, we can remove the first element of $\thinList(\rooted{T_r}{v})$ to obtain the thinness list of $\rooted{T'_r}{v}$. Again, we store these values in new variables to be able to restore them later.

        It remains for us to obtain the information on the child subtrees of $v$ in $T'_r$ in the last case. There is only one child $g$ of $v$ in $T_r$ such that $\rooted{T_r}{g}$ has thinness $k$ and has a critical vertex, as otherwise $\rooted{T_r}{v}$ would have thinness at least $k+1$ by \cref{lemma:one-critical-vertex}. We thus proceed in the same way as we did to get the information on $\rooted{T'_r}{v}$. We keep all information on the child subtrees of $v$ different from $\rooted{T_r}{g}$ without modification. If a child of $g$ is the critical vertex, we remove the information on $\rooted{T_r}{g}$ from $\gchildInfo$. If not, we remove the first entry in $\critList(\rooted{T_r}{g})$ and $\thinList(\rooted{T_r}{g})$ to obtain $\critList(\rooted{T'_r}{g})$ and $\thinList(\rooted{T'_r}{g})$, respectively.

        To summarize:
        \begin{itemize}
            \item The information for the child subtrees of $r$ in $T'_r$ different from $\rooted{T'_r}{v}$ remains the same in the recursive call.
            \item If a child of $v$ is the critical vertex in $\rooted{T_r}{v}$ then we remove the information from $\childInfo$ and $\gchildInfo$ corresponding to $\rooted{T_r}{v}$ and the child subtrees of $v$ in $T_r$.
            \item Otherwise, we remove the first elements of $\critList(\rooted{T_r}{v})$ and $\thinList(\rooted{T_r}{v})$ to obtain $\critList(\rooted{T'_r}{v})$ and $\thinList(\rooted{T'_r}{v})$, respectively. We iterate over all the child subtrees of $v$ in $\gchildInfo$ to see which one has thinness $k$ and has a critical vertex. This can be done by looking at the first elements of their thinness lists and critical vertex lists. Once we find a child $g$ of $v$ that satisfies this, we do not change any other information on the child subtrees of $v$ on the recursive call, and we either remove the information on $\rooted{T_r}{g}$ if a child of $g$ is a critical vertex, or we remove the first element of $\critList(\rooted{T_r}{g})$ and $\thinList(\rooted{T_r}{g})$ to obtain $\critList(\rooted{T'_r}{g})$ and $\thinList(\rooted{T'_r}{g})$, respectively.
        \end{itemize}
            
        After obtaining the information on the child and grandchild subtrees of $T'_r$, we call \textproc{ComputeThinnessAndCritLists}, and use the first element of the thinness list of $T'_r$ to check if Condition 6 is met or not. 

        All the steps necessary to obtain the information on $T'_r$ can be accomplished by iterating over the elements of $\childInfo$ and $\gchildInfo$, which means they take $\bigO(\abs{\child(r)} + \abs{\gchild(r)})$ time. However, this is not enough to determine the running time of Condition 6, as we are calling \textproc{ComputeThinnessAndCritLists} recursively, and have not yet analyzed the running time of the recursive call.

        Recall that $\rooted{T_r}{v}$ is the only child subtree of $r$ in $T_r$ with thinness $k$. In other words, $\rooted{T_r}{v}$ is the child subtree that has $k$ as the first element of its thinness list. To obtain the arguments passed to the recursive call, we either removed the thinness list of $\rooted{T_r}{v}$ entirely from $\childInfo$, or we removed its first element. This fact combined with \cref{lemma:thinness-of-Tri-is-bigger-than-thinness-of-Tri+1} implies that the greatest element of any thinness list in $\childInfo$, if it is not empty, is strictly less than $k$. In general, in every recursive call to \textproc{ComputeThinnessAndCritLists}, the greatest element of any thinness list in $\childInfo$ will be reduced if it is not empty, as the only time we make a recursive call is in the processing of Condition 6. Furthermore, no subtree can have thinness lower than 1. If $\childInfo$ is empty, $r$ is a leaf, and thus we reach a base case. Therefore, the depth of the recursion is bounded by $k$, which in turn is bounded by the thinness of a child subtree of $r$ in $T_r$. By \cref{log-bound}, this bound is $\bigO(\log(n))$. Every other computation necessary in a recursive call takes $\bigO(\abs{\child(r)} + \abs{\gchild(r)})$, and the amount of children and grandchildren of $r$ does not increase with each successive call. Therefore, the total running time of the recursive call is $\bigO((\abs{\child(r)} + \abs{\gchild(r)}) \cdot \log(n))$, which is the total running time of computing Condition 6.
    \end{itemize}
    
    We have described how to check the conditions in the decision tree of \cref{fig:decision-tree} in the runtime complexity required by the statement. Now, we describe how to compute the critical vertex list and the thinness list of $T_r$ in each case.
    
    It was shown in \cref{lemma:decision-tree} that the thinness of the tree is correctly computed in each case. This will be, by definition, the first element of the thinness list. We therefore show how to compute the critical vertex list and the rest of the thinness list.

    First, note that, if $\thin(T_r) = k+1$, then $T_r$ has no critical vertex, as none of its complete rooted subtrees have thinness $k+1$. Therefore, the critical vertex list of $T_r$ in these cases is $(\nil)$, and its thinness list is $(k+1)$.

    It remains to analyze the cases where $\thin(T_r) \neq k+1$.
    
    \begin{itemize}
        \item \emph{Case 1}: In this case, $r$ is a leaf. Tree $T_r$ has no critical vertices, and therefore its critical vertex list is $(\nil)$, and its thinness list is $(k)$.
        \item \emph{Case 3}: There are no child subtrees of $r$ with thinness $k$ that contain a critical vertex. Thus, the only possible critical vertex of $T_r$ is $r$. If $\abs*{\N[T_r]{k}{r}} = 2$, then $\critList(T_r) = (r)$, otherwise, $\critList(T_r) = (\nil)$. In both cases, the thinness list is $(k)$, as by the \defref{def:thinness-list} the length of the thinness list is the same as the length of the critical vertex list. These can be computed in constant time having computed $\abs*{\N[T_r]{k}{r}}$ for Condition 2.
        \item \emph{Case 5}: In this case, a child subtree $\rooted{T_r}{v}$ of $r$ with thinness $k$ has a critical vertex as its root. Its critical vertex list is $(v)$, and its thinness list is $(k)$.
        \item \emph{Case 7}: In this last case, the critical vertex $x$ of $T_r$ is neither $r$ nor a child of $r$. By the \defref{def:critical-vertex-list}, its critical vertex list is the result of adding $x$ to the beginning of the critical vertex list of $T_r \setminus \rooted{T_r}{w}$, and by \cref{obs:suffix-of-thinness-list-is-thinness-list-of-Tr2}, its thinness list is the result of adding $k$ to the beginning of the thinness list of $T_r \setminus \rooted{T_r}{w}$.
        We can compute the critical vertex list and the thinness list of $T_r \setminus \rooted{T_r}{w}$ in the recursive call to \textproc{ComputeThinnessAndCritLists} we performed when analyzing Condition 6. This can be done in a runtime which is proportional to the size of the thinness list of $T_r \setminus \rooted{T_r}{w}$, which is $\bigO(\log(n))$ by \cref{lemma:thinness-list-has-log-n-elements}.
    \end{itemize}

    All these steps, except the ones in Case 7, can be performed in constant runtime.
    
    Combining all the running times necessary to check the conditions and assign the correct critical vertex list and thinness list results in a runtime complexity of $\bigO((\abs{\child(r)} + \abs{\gchild(r)}) \cdot \log(n))$, as the statement claims.
\end{proof}

Now, we describe how to apply the procedure \textproc{ComputeThinnessAndCritLists} on each of the vertices of a tree to compute its thinness.

\begin{theorem}
    \label{thm:thinness-algorithm}
    Given a tree $T$ on $n$ vertices, $\thin(T)$ can be computed in $\bigO(n\log(n))$-time.
\end{theorem}

\begin{proof}
We assign a root $r$ to $T$ arbitrarily to obtain $T_r$, and call the procedure \textproc{ComputeThinnessAndCritLists} defined in \cref{lemma:compute-list-and-type} with tree $T_r$ and the critical vertex lists and the thinness lists of the child and grandchild subtrees of $r$ as argument. Part of the output will be the thinness list of $T_r$, which will have the thinness of $T$ as its first element.

To obtain the necessary critical vertex lists and thinness lists, we start by calling \textproc{ComputeThinnessAndCritLists} on the leaves of $T_r$. As these vertices have no children, arguments $\childInfo$ and $\gchildInfo$ will just be two empty lists for these calls. Then, we proceed by using the outputs of these calls to call \textproc{ComputeThinnessAndCritLists} on the parents of the leaves, and advance in a bottom-up fashion until we reach $r$ to compute the critical vertex lists and the thinness lists of all the complete rooted subtrees of $T_r$. 

This algorithm performs a call to \textproc{ComputeThinnessAndCritLists} for each rooted complete subtree of $T_r$. There are exactly $n$ rooted complete subtrees of $T_r$; one for each vertex in $T_r$. By \cref{lemma:compute-list-and-type}, the call with argument $\rooted{T_r}{v}$ takes time proportional to $\bigO((\abs{\child(v)} + \abs{\gchild(v)}) \cdot \log(n))$, so the total running time is \[\bigO\left(\sum_{v \in V(T_r)}(\abs{\child(v)} + \abs{\gchild(v)}) \cdot \log(n)\right).\]

Every vertex is the child of at most one vertex, and it is the grandchild of at most one other vertex. Therefore, \[\sum_{v \in V(T_r)}(\abs{\child(v)} + \abs{\gchild(v)}) \leq 2n.\]
Then, the total running time of the algorithm is $\bigO(2n\log(n))$, which is equal to $\bigO(n\log(n))$.
\end{proof}

\subsection{Finding an optimal consistent solution}

We conclude with an algorithm that uses the information produced by the algorithm described in \cref{thm:thinness-algorithm} to find an optimal consistent solution for $T$ in the same time complexity.

\begin{theorem}
    A consistent solution using $\thin(T)$ classes for a tree $T$ on $n$ vertices can be found in $\bigO(n\log(n))$-time.
\end{theorem}
\begin{proof}
    First, we run the algorithm described in \cref{thm:thinness-algorithm} on $T$, which computes the critical vertex lists and the thinness lists of all the complete rooted subtrees of $T$ as part of its operation by calling \textproc{ComputeThinnessAndCritLists} repeatedly. We store this information in a table $A$ indexed by the root of the complete rooted subtree so that we can access it in constant time for any complete rooted subtree of $T$.

    We define the procedure \textproc{ConsistentSolution} that, given a rooted tree $T_r$ and a table $A$ with the corresponding information for all complete rooted subtrees of $T_r$, will:
    \begin{enumerate}
        \item Construct a path $P$ such that all the connected components of $T_r \setminus N[P]$ have thinness lower than $T_r$.
        \item Recursively compute a consistent solution for those components.
        \item Apply the \nameref{lem:path-layout} to obtain a consistent solution for $T$. 
    \end{enumerate}

    Let $k$ be the thinness of $T_r$. Suppose that $T_r$ has no critical vertex. In other words, the first element of $\critList(T_r)$ is $\nil$. For any vertex $v \in V(T_r)$, at most one of the children of $v$ is a $k$-neighbor of $v$. Denote it $w$ if it exists. We can identify $w$ by iterating over all the grandchild subtrees of $v$ to see which one has $k$ as the first element of its thinness list. We take $P$ to be the longest possible path of the form $(v_0, v_1, ..., v_{\text{last}})$, where $\text{last}\in \NN$, $v_0=r$, and $v_i$ is the only $k$-neighbor that is a child of $v_{i-1}$ for all $i \in \{1, \dots, \text{last}\}$. Notice that this path is unique. All the parents of the vertices in $P$ are also in $P$, and thus all connected components of $T \setminus P$ are complete rooted subtrees of $T_r$, as are all connected components of $T \setminus N[P]$. Also, no vertex in $P$ has a child $k$-neighbor outside $P$, which means that no child of a vertex in $P$ has itself a child that is the root of a subtree with thinness $k$. Therefore, $T \setminus N[P]$ has no connected component with thinness greater than $k - 1$. If $T \setminus N[P]$ is empty, this is a base case, and no further processing is needed. Otherwise, we call \textproc{ConsistentSolution} recursively on the subtrees in $T \setminus N[P]$ to compute a consistent solution for each one and then apply the procedure \textproc{ConsistentSolutionGivenPath} described in the \nameref{lem:path-layout} to obtain a consistent solution for $T$.

    Now, suppose that $x$ is the first element of $\critList(T_r)$. There are exactly two child $k$-neighbors $v_1$ and $v_2$ of $x$. By \cref{lemma:one-critical-vertex}, $\rooted{T_r}{v_1}$ and $\rooted{T_r}{v_2}$ have no critical vertices. As seen before, for each $i \in \{1,2\}$ we can thus build a path $P_i$ that has no $k$-neighbors in $\rooted{T_r}{v_i}$ that starts at $v_i$ and ends at a descendant $w_i$ of $v_i$. We take $P$ to be the path  $(w_1, \dots, v_1, x, v_2, \dots, w_2)$. Path $P$ has no $k$-neighbors in $\rooted{T_r}{v_1}$ and $\rooted{T_r}{v_2}$. Additionally, there are no $k$-neighbors of $x$ other than $v_1$ and $v_2$, and thus $P$ has no $k$-neighbors in any other subtree of $T_r$. Therefore, all connected components in $T \setminus N[P]$ have thinness lower than $k$.

    In this case, however, the connected components of $T \setminus N[P]$ are not necessarily all complete rooted subtrees of $T_r$. In particular, if $x$ has a parent $p$ and a grandparent $h$, then $\dangling{T_r}{p}{h}$ is not a complete rooted subtree of $T_r$. We will need to update the elements of table $A$ so that it contains the information on the complete rooted subtrees of $\dangling{T_r}{p}{h}$ when calling \textproc{ConsistentSolution} recursively.
    
    Let $T'_r = \dangling{T_r}{p}{h}$. Let $y$ be an ancestor of $x$ in $T_r$ different from $p$. By \cref{obs:critical-vertex-of-parent-is-same-as-descendant}, the critical vertex of $\rooted{T_r}{y}$ is also $x$. With this we see, by the \defref{def:critical-vertex-list}, that the critical vertex list of $\rooted{T'_r}{y}$ is the suffix of $\critList(\rooted{T_r}{y})$ starting at position 2. Similarly, the thinness list of $\rooted{T'_r}{y}$ is the suffix of $\thinList(\rooted{T_r}{y})$ starting at position 2. We can thus remove the first elements of these two lists in table $A$, and $A$ will contain the information on $\rooted{T'_r}{y}$. We can repeat this process for every ancestor $y$ of $x$ in $T_r$ different from $p$ to obtain the updated information on all the subtrees rooted on them.

    \Vref{fig:path-construction} illustrates the construction of path $P$ following the approach described.    
    \newcommand{\isotriangle}[2][10mm]{
        \node[draw=black, shape border rotate=90, isosceles triangle, minimum height=#1, anchor=north] at (#2) {};
    }
    \tikzstyle{nodo}=[inner sep=1, draw, circle, fill=black]
    \tikzstyle{simplepath}=[rounded corners, opacity=0.2, draw=blue, line width=5mm, line cap=round]
    \begin{figure}[htbp]
        \centering
        \begin{subfigure}{0.4\textwidth}
        \centering
        \begin{tikzpicture}[sibling distance=10mm]
            \node[nodo, label=north:{$v_0=r$}] (v0) {}
                child {
                    coordinate (t1)
                }
                child {
                    coordinate (t2)
                }
                child {
                    node[nodo, label=east:$v_1$] (v1) {}
                    child {
                        node[nodo, label=east:$v_2$] (v2) {}
                        child {
                            node[nodo, label=west:$v_3$] (v3) {}
                            child {
                                coordinate (t4)
                            }
                            child {
                                coordinate (t5)
                            }
                        }
                        child {
                            coordinate (t3)
                        }
                    }
                }
                ;
            
            \foreach \t in {1,...,5} {\isotriangle{t\t}}
            \path[simplepath, line width=8mm] (v0.center) -- (v1.center) -- (v2.center) -- (v3.south);
        \end{tikzpicture}
        \end{subfigure}
        \begin{subfigure}{0.4\textwidth}
            \centering
            \begin{tikzpicture}[sibling distance=10mm]
                \node[nodo, label=north:$r$] (v0) {}
                child {
                    coordinate (t1)
                }
                child {
                    node[nodo, label=west:$h$] (v1) {}
                    child {
                        node[nodo, label=west:$p$] (v2) {}
                        child {
                            node[nodo, label=west:$x$] (v3) {}
                            child {
                                node[nodo, label=west:$v_1$] (v4) {}
                                child {
                                    node[nodo] (v5) {}
                                    child {
                                        coordinate (t6)
                                    }
                                    child {
                                        node[nodo, label=east:$w_1$] (v6) {}
                                        child {
                                            coordinate (t7)
                                        }
                                    }
                                }
                                child {
                                    coordinate (t4)
                                }
                            }
                            child {
                                coordinate (t8)
                            }
                            child {
                                node[nodo, xshift=3mm, label=east:$v_2$] (v7) {}
                                child {
                                    node[nodo, label=west:$w_2$] (v8) {}
                                }
                                child {
                                    coordinate (t5)
                                }
                            }
                        }
                        child {
                            node[xshift=10mm, inner sep=0] (t3) {}
                        }
                    }
                }
                child {
                    coordinate (t2)
                }
                ;
            
            \foreach \t in {1,...,8} {\isotriangle{t\t}}
            
            \path[simplepath, line width=8mm] (v6.center) -- (v5.center) -- (v4.center) -- (v3.center) -- (v7.center) -- (v8.center);
            \path[simplepath, draw=red] (v4.center) -- (v5.center) -- (v6.center);
            \path[simplepath, draw=red] (v7.center) -- (v8.center);    
            \end{tikzpicture}
        \end{subfigure}
        \caption{Construction of path $P$ depending on the existence of a critical vertex. Path $P$ is colored in blue. On the left, the tree has no critical vertex. Here, $v_{i+1}$ is the child $k$-neighbor of $v_i$ for each $i \in \{0,1,2\}$. On the right, the tree has a critical vertex $x$, and $v_1$ and $v_2$ are the child $k$-neighbors of $x$.}
        \label{fig:path-construction}
    \end{figure}
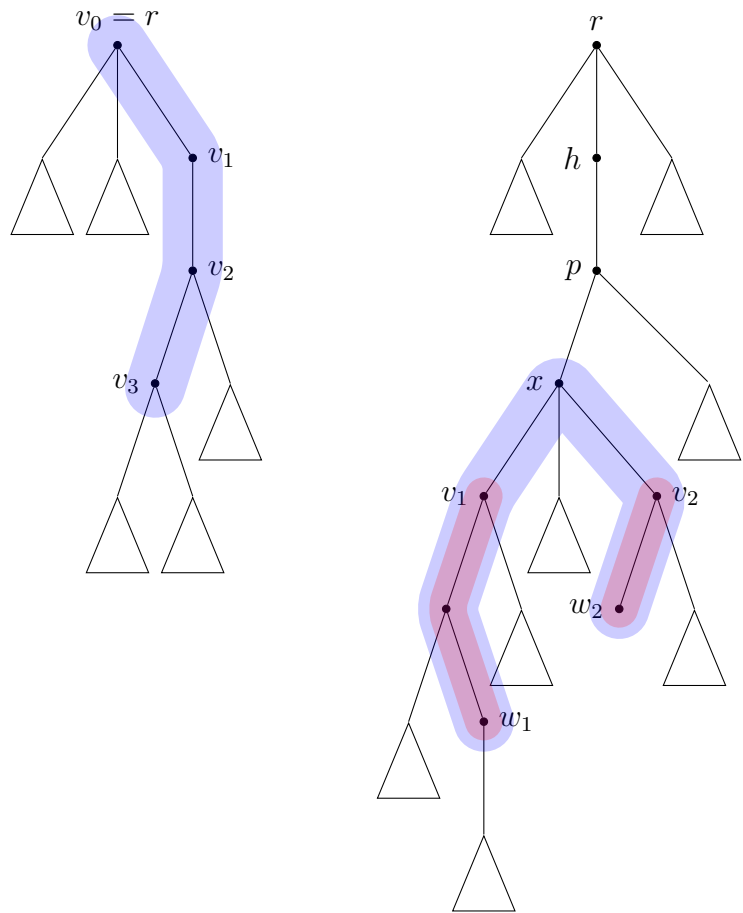

    Note that every recursive call is made with a strict subtree of $T$ as argument, and thus the algorithm has recursion depth at most $n$.

    Some implementation details must be detailed in order to reach the desired time complexity $\bigO(n \log(n))$. The order computed by \textproc{ConsistentSolution} is represented by a linked list of vertices, which has constant time complexity when adding or removing a vertex before or after another vertex, and concatenating two lists. The partition is represented by a list of linked lists. Each linked list represents a class in the partition, and contains the vertices corresponding to that class. These data structures are exactly the ones returned by \textproc{ConsistentSolutionGivenPath}. The paths and the neighborhoods of the paths will also be represented with linked lists, as they will be the actual classes in the partition. Notice that the size of the partition is $\bigO(\log(n))$ by \cref{log-bound}. 

    Now, we analyze the time complexity of this approach. The operations performed by the algorithm are:
    \begin{itemize}
        \item \emph{Adding a vertex to a path.} This operation is performed a number of times proportional to $n$, as each vertex is added to a single path. Also, this operation has constant time complexity, as it consists of simply adding an element to the start or the end of a linked list.
        \item \emph{Finding the child $k$-neighbor of a vertex $v$.} This operation is performed a number of times proportional to $n$, as this is only performed right after adding $v$ to $P$. This operation has a time complexity proportional to the number of grandchildren of $v$. As the sum of grandchildren over all vertices is proportional to $n$, the time spent on operations of this type is $\bigO(n)$.
        \item \emph{Removing an entry from $A$.} The number of entries of $A$ is equal to $n$, so this operation is performed a number of times proportional to $n$. This operation has constant time complexity, as it consists of simply marking an entry of a table as empty.
        \item \emph{Updating an entry in $A$.} The update operation involves removing the first element of two lists, and thus is of constant time complexity. The thinness list has at most $\bigO(\log(n))$ entries by \cref{lemma:thinness-list-has-log-n-elements}, and thus a specific entry cannot be updated more than $\bigO(\log(n))$ times. As there are $n$ entries in $A$, the time spent on operations of this type is $\bigO(n \log(n))$.
        \item \emph{Combining the results of recursive calls to \textproc{ConsistentSolution} by calling \textproc{ConsistentSolutionGivenPath}.}
        As described in \cref{lem:path-layout}, this operation has time complexity $\bigO(\abs{V(N[P])} + k \cdot q)$, where $P$ is the path constructed by our algorithm, $k + 1$ is the thinness of the subtree $T'$ of the tree $T$, and $q$ is the number of connected components in $T' \setminus N[P]$. First, each vertex of the original tree $T$ does not appear twice in the neighborhood of a path as an argument in a call to \textproc{ConsistentSolutionGivenPath}, because each recursive call to \textproc{ConsistentSolution} receives as argument a subtree of $T$ that does not contain any vertex already assigned to a class. Second, $k$ can be bounded by $\log(n)$ by \cref{log-bound}. Lastly, the sum of the values of $q$ over all recursive calls is proportional to the number of edges in $T$, which is $n - 1$. This is because each connected component of $T' \setminus N[P]$ is determined by the edge that connects the vertex $v$ in $N[P]$ to the connected component in $T'$, and $v$ appears in just one neighborhood of a path in the whole recursive procedure.Therefore, the time spent on operations of this type is $\bigO(n + \log(n) \cdot (n - 1)) = \bigO(n \log(n))$.
    \end{itemize}
    
    The rest of the operations performed by the algorithm take constant time. Therefore, the total time complexity of the algorithm is $\bigO(n \log(n))$.
\end{proof}

\section*{Acknowledgements}

This work was partially supported by CONICET (PIP 11220200100084CO), ANPCyT (PICT-2021-I-A-00755) and UBACyT (20020220300079BA and 20020190100126BA).

%
%

\end{document}